\documentclass[
reprint,
 amsmath,amssymb,aps,
doublecolumn,
floatfix,longbibliography
]{revtex4-2}

\usepackage{graphicx}
\usepackage{dcolumn}
\usepackage{bm}
\graphicspath{{figs/}}

 \usepackage{xcolor} 
\begin{document}

\preprint{APS/123-QED}

\title{One dimensional cell motility patterns}

\author{Jonathan E. Ron$^{1}$, Pascale Monzo$^{2}$, Nils C. Gauthier$^{2}$, Raphael Voituriez$^{3}$ and Nir S. Gov$^{1}$}
\affiliation{$^{1}$Department of Chemical and Biological Physics, Weizmann Institute of Science, Israel}
\affiliation{$^{2}$IFOM, FIRC Institute of Molecular Oncology, Milan, 20139, Italy}
\affiliation{$^{3}$Laboratoire Jean Perrin and Laboratoire de Physique Théorique de la Matière Condensée, Sorbonne Université, Tour 13-12, 5eme etage, 4 place Jussieu, 75252 Paris Cedex 05, France}

\begin{abstract}
During migration cells exhibit a rich variety of seemingly random migration patterns, which makes unraveling the underlying mechanisms that control cell migration a daunting challenge. For efficient migration cells require a mechanism for polarization, so that traction forces are produced in the direction of motion, while adhesion is released to allow forward migration. To simplify the study of this process cells have been studied when placed along one-dimensional tracks, where single cells exhibit both smooth and stick-slip migration modes. The stick-slip motility mode is characterized by protrusive motion at the cell front, coupled with slow cell elongation, which is followed by rapid retractions of the cell back. In this study, we explore a minimal physical model that couples the force applied on the adhesion bonds to the length variations of the cell and the traction forces applied by the polarized actin retrograde flow. We show that the rich spectrum of cell migration patterns emerges from this model as different \emph{deterministic} dynamical phases. This result suggests a source for the large cell-to-cell variability (CCV) in cell migration patterns observed in single cells over time and within cell populations. The large heterogeneity can arise from small fluctuations in the cellular components that are greatly amplified due to moving the cells' internal state across the dynamical phase transition lines. Temporal noise is shown to drive random changes in the cellular polarization direction, which is enhanced during the stick-slip migration mode.
These results offer a new framework to explain experimental observations of migrating cells, resulting from noisy switching between underlying deterministic migration modes.
\end{abstract}

\maketitle


\section{\label{sec:intro}Introduction}
Eukaryote cell migration, whereby cells crawl actively over an external substrate, is a subject of great interest for biological processes such as development and cancer progression. Adhesion-based cell motility involves the orchestration of a large number of cytoskeletal proteins: Typically, the cell needs to break its symmetry (polarize), and produce traction forces in the direction of polarization (balanced by drag or friction forces). The traction forces are mediated by adhesion between the cell and the external substrate, but these adhesions have to detach at the trailing end of the cell in order to allow the cell to migrate forward. When observing freely migrating cells (i.e. not guided by an external gradient of any kind), one is often struck by the large variation in their migration patterns \cite{shreiber2003temporal,keren2008mechanism,takagi2008functional,wu2014three,kowalewski2015disentangling}, both for a single cell over time, and for a (seemingly identical) cell population. The origin of the large Cell-to-cell variability (CCV), or phenotypic, population heterogeneity, observed during cell migration is not understood \cite{bruckner2019disentangling}, and is usually ascribed to the inherent noise of cellular systems \cite{petrie2009random,miao2017altering}. 

To simplify the study of the complex process of cell motility, cells can be confined to move along one-dimensional tracks, either on flat adhesive stripes \cite{monzo2016mechanical,hennig2020stick}, linear grooves \cite{han2013unidirectional} and channels \cite{stroka2014water,chabaud2015cell}, or thin fiber \cite{sheets2013shape,guetta2015protrusive}. In addition to being a simple geometry for the study and analysis of the cell motion in general, such confined motion appears also in-vivo \cite{friedl2009collective}, for example when cells move along axonal fibers \cite{rakic1990principles,parnavelas2000origin}, or cancer invades in confined spaces between tissues \cite{paul2017cancer}.

The experiments listed above have shown that isolated cells on one-dimensional tracks exhibit the following stereotypical behaviors \cite{monzo2016mechanical,hennig2020stick}: (i) Non-migrating and unpolarized, by remaining quiescent or elongating symmetrically, (ii) Undergoing spontaneous symmetry breaking, polarization and migrating smoothly, (iii) As in (ii) but exhibiting stick-slip migration. The stick-slip motility mode is characterized by protrusive motion at the cell front, coupled with an overall elongation of the cell and followed by rapid retraction of the cell back. In both (ii,iii) the cell motility can be highly persistent, or undergo sporadic direction changes. The appearance of this variety of migration behaviors within a uniform cell population, as well as the switching of cells between these modes, remains an open puzzle and is the aim of this work. Clearly the diversity of the migration patterns is beyond modeling the cell motility as a simple random walk process \cite{danuser2013mathematical,wu2014three,selmeczi157s,vestergaard2015estimation}.

We present a theoretical model that describes the cell motility in a highly simplified, and coarse-grained manner. Nevertheless, the model contains two key, and strongly coupled, components: a slip-bond adhesion module at the cell back, and a cellular polarization module. We find that these components are sufficient to drive the entire spectrum of observed motility patterns, and explain the transitions between them. Our work therefore demonstrates how a minimal model gives rise to a rich variety of \emph{deterministic migration patterns}. In experiments these \emph{deterministic} patterns may underlie the migration of cells, but further confounded by noise.

The stick-slip motion is shown here to have an underlying deterministic oscillatory behavior, separated from smooth migration by a bifurcation line. Unlike previous treatments of stick-slip dynamics of cells, which focused on the protrusion-retraction cycles at the cell edge \cite{chan2008traction,sens2013rigidity}, we couple the stick-slip adhesion at the cell back to the overall cell polarization, through the dependence of the polarization on the cell length. This dependence, which is an inherent property of the UCSP model \cite{maiuri2015actin}, was not previously explored. Deterministic oscillations in the speed of migrating dendritic cells, for example, were related to competition for finite resources that directly affected the acto-myosin polarization mechanism, but did not involve length oscillations \cite{lavi2016deterministic}. Furthermore, the dendritic-cell oscillations depend on a specific macropinocytosis process, while here we obtain deterministic oscillatory (stick-slip) migration patterns that are driven by adhesion dynamics that are much more general across cell types. 

\section{\label{sec:model}Model}
The model is introduced by three parts, with increasing levels of complexity and realism. In this manner we expose the motility patterns and the key components that drive them.

The first part describes a cell that is constantly polarized, with a constant protrusive activity at the leading edge, and slip bond adhesions at the rear \cite{sens2013rigidity}. This part allows us to expose the oscillatory stick slip behavior through the dynamics of the cell length and adhesion concentration at the rear.

The second part adds a self-polarization model to the polarized cell \cite{maiuri2015actin}. This part couples the dynamics in cell length to the protrusive activity, and introduced a critical polarization length scale. However the model simplifies the cell to have a single leading edge, where all the protrusive activity is concentrated.

In the third part the model is extended to be symmetric, such that the protrusion and adhesion dynamics acts on both edges of the cell. This part outlines the conditions for symmetry breaking and the role of noise in choosing a migration direction.

\subsection*{\label{sec:model0}Part 1: Polarized cell with a constant protrusion}
\subsubsection*{Model description}
\begin{figure}[htbp!]
\includegraphics[width=\linewidth]{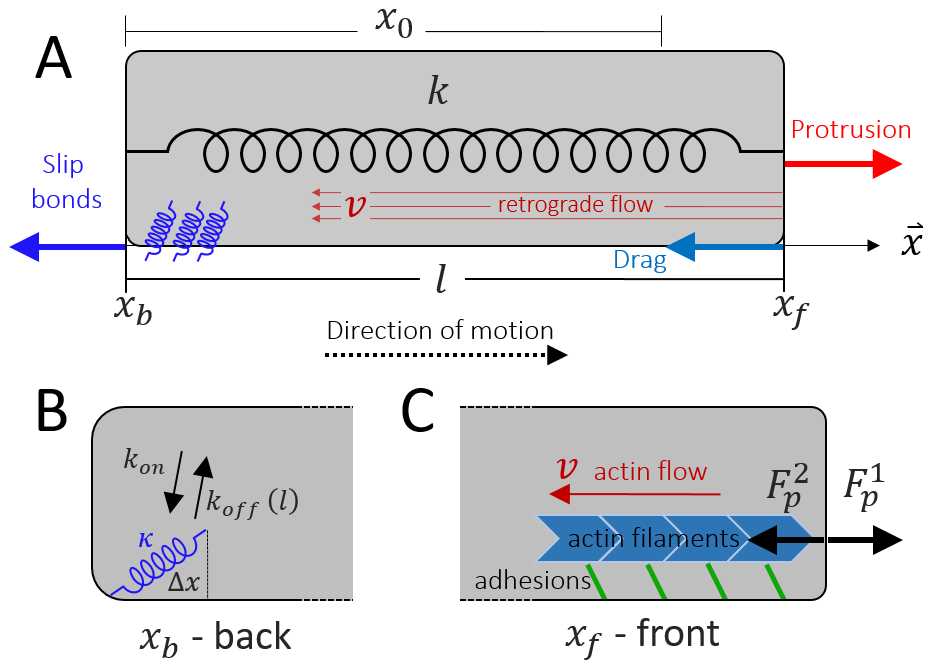}
\caption{\label{fig:constV_01} The simplified model. A) Illustration of a the physical model of a cell migrating along a linear track. $x_b$ and $x_f$ represent the back and front part of the cell of the cell which are connected by a spring with a stiffness $k$. $x_0$ is the rest length of the spring and $l$ is the length of the cell. $v$ is the velocity of the actin retrograde flow which is assumed to be constant. At the front acts a protrusion force (red arrow) and a drag force (teal arrow). At the back acts a friction force due to the slip bonds (blue arrow). B) The physical model of the stick slip adhesion at the back $x_b$. Stochastic linkers with stiffness $\kappa$ attach with an average rate of $k_on$ and detach with a length dependent rate of $k_off(l)$ (Eq.\ref{koff0}) at the back. The linkers stretch on average with a displacement of $\Delta x$ (Eq.\ref{Dx}). C) The physical model of the protrusion force acting at the front $x_f$: $F_p^{1}$ and $F_p^2$ (Eqs.\ref{Fp1},\ref{Fp2}) are forces that act on/from the sliding actin filaments, which are balanced by catch adhesions (green short lines).}
\end{figure}

Consider a cell of length $l$ that migrates along a linear track. The two ends of the cell, the front and back, denoted by $x_f$ and $x_b$, are connected by a spring (Fig. \ref{fig:constV_01}A). The stiffness of the spring $k$, represents the effective elasticity of the cell cytoplasm and membrane. Such a mechanical coupling between the front and rear was recently demonstrated experimentally \cite{tsai2019efficient}.

In this part of the model, the motion of the cell is considered to be already polarized, such that the actin treadmilling from the front to the back occurs at a constant velocity $v$, and produces a constant protrusive force at the front
\begin{equation} \label{Fp1}
    F_p^{1}=\alpha v
\end{equation}
where $\alpha$ describes the strength of the coupling of the actin retrograde flow and the effective friction force generated by adhesions which grip the sliding filaments. This is a simplified representation of the "clutch" mechanism \cite{chan2008traction}, that converts the sliding of the actin to a protrusive force that pushes on the membrane. This coupling is dependent on the adhesion strength, and we therefore expect a term of the form: $r/(r+r_0)$ to multiply the r.h.s. of Eq.\ref{Fp1}, where $r$ is the ratio between the binding/unbinding rates of the cell-substrate adhesion molecules $r=k_{on}/k_{off}^0$, and $r_0$ quantifies the the cell-substrate adhesion saturation. Such a term accounts for the loss of traction force when the adhesion diminishes $r\rightarrow 0$. We do not explicitly describe here the catch-bond property of these adhesions, as we are not interested in the stick-slip dynamics of the leading edge, but rather wish to focus on the stick-slip events on the whole cell scale. For the rest of this paper we effectively work in the limit of  $r_0\ll r$ (the effects of $r_0$ are shown in Fig.\ref{fig:SI_02}).

The pushing force is balanced by a drag force which is proportional to the speed of the moving front, and a restoring force due to the global cell elasticity
\begin{equation} \label{Fp2}
    F_{p}^2=\gamma\dot{x}_f+k(x_f-x_b-x_0)
\end{equation}
where $\gamma$ represents the effective resistance to the motion of the cell front due to the friction generated by the contact of adhesion molecules with the substrate. As in Eq.(\ref{Fp1}), we expect a term of the form $r/(r+r_0)$ to multiply the first term on the r.h.s. of Eq.\ref{Fp2}, but we neglect it by choosing to work in the limit of $r_0\ll r$. The parameter $k$ represents the stiffness of the spring which describes the effective elasticity of the cell, and the parameter $x_0$ represents the rest length of the cell.

Equating (\ref{Fp1}) and (\ref{Fp2}) due to the force balance at $x_f$ yields
\begin{equation} \label{Fp1=Fp2}
\alpha v = \gamma \dot{x}_f+k(x_f-x_b-x_0)
\end{equation}
which allow to obtain the equation of motion for the moving cell front
\begin{equation} \label{xf}
    \dot{x}_f=\frac{1}{\gamma}\left(\alpha v-k\left(x_f-x_b-x_0\right)\right)
\end{equation}
At the back, $x_b$, the pulling force due to the cell elasticity
\begin{equation} \label{Fb1}
    F_{back}^+ = k(x_f-x_b-x_0) 
\end{equation}
is balanced by a friction force which results from the stretching of bound linkers, which model slip bond adhesions 
\begin{equation} \label{Fb2}
    F_{back}^- = n\kappa \langle\Delta x\rangle
\end{equation}
where $n$ is the mean number of bound linkers, $\kappa$ is the spring constant of the linkers, and $\langle\Delta x\rangle$ is the average displacement of the linkers (Fig.\ref{fig:constV_01}C).
The number of bound linkers evolves dynamically and obeys the following kinetics
\begin{equation} \label{nDot}
    \dot{n}=k_{on}(N-n)-k_{off}n
\end{equation}
where $N$ is the total number of linkers, $k_{on}$ is the basal attachment rate, and $k_{off}$ is a detachment rate which depends exponentially on the force applied on each bound linker  \cite{sens2013rigidity,chan2008traction}
\begin{equation} \label{koff0}
    k_{off}=k_{off}^0 exp\left(\frac{f_l}{f_s}\right)
\end{equation}
where $k_{off}^0$ is the basal detachment rate and $f_s$ represents the susceptibility of the linkers to the applied force \cite{sens2013rigidity}. The stretching force per bound linker is
\begin{equation}\label{fl}
    f_l=\frac{k(x_f-x_b-x_0)}{n}
\end{equation}
The average stretch of each linker $\langle \Delta x \rangle$ depends on the dissociation rate of the linkers (\ref{koff0},\ref{nDot}) and can be written as
\begin{equation} \label{Dx}
    \langle \Delta x \rangle = \frac{{\dot{x}}_b}{k_{off}}
\end{equation}
By combining (\ref{koff0},\ref{fl},\ref{Dx}) into the force balance between (\ref{Fb1}) and (\ref{Fb2}) we obtain
\begin{equation} \label{kappa}
    n \kappa \frac{\dot{x}_b}{k_{off}^0 exp\left(\frac{k(x_f-x_b-x_0)}{n f_s}\right)}=k\left(x_f-x_b-x_0\right)
\end{equation}
where $\kappa$ is the effective spring constant of the linkers, and we substituted for the average stretching of each linker: $\langle\Delta x\rangle=\dot{x}_b/k_{off}$. 
By reorganizing (\ref{kappa}) we obtain the equation of motion of the moving back 
\begin{equation} \label{xb}
    \dot{x}_b = \frac{k(x_f-x_b-x_0)}{n \kappa}k_{off}^0 exp\left(\frac{k(x_f-x_b-x_0)}{n f_s}\right)
\end{equation}

Combining (\ref{xf}) and (\ref{xb}), and changing coordinates to $l=x_f-x_b$ yields the following dynamical system
\begin{align} \label{ds1A}
    \dot{l}=&\frac{\alpha}{\gamma}v-k(l-x_0)\left(\frac{1}{\gamma}-\frac{k_{off}^0}{n \kappa} exp\left(\frac{k(l-x_0)}{n f_s}\right)\right)\\ \label{ds1B}
    \dot{n}=&k_{on}(N-n)-k_{off}n
\end{align}
Next, the system (\ref{ds1A},\ref{ds1B}) is rescaled by the time and length scales of $k_{off}^{-1}$,$x_0$, and by the total number of adhesion sites $N$, as well as 
rescaled by the parameters $\frac{v}{x_0k_{off}^0}\rightarrow \tilde{v}$, $\frac{k}{k_{off}^0}\rightarrow \tilde{k}$, $\frac{f_s N}{x_0 k_{off}^0}\rightarrow \tilde{f}_s$, $\frac{N\kappa}{k_{off}^0}\rightarrow \tilde{\kappa}$ and $\frac{k_{on}^0}{k_{off}^0}\rightarrow r$ to obtain
\begin{align} \label{ds3A}
    \dot{l}=&\frac{\alpha}{\gamma} \tilde{v}-\tilde{k}(l-1)\left(\frac{1}{\gamma}-\frac{1}{\tilde{\kappa} n}exp\left(\frac{\tilde{k}(l-1)}{\tilde{f}_s n}\right)\right)\\ \label{ds3B}
    \dot{n} =& r(1-n)-n\cdot exp\left(\frac{\tilde{k} (l-1)}{\tilde{f}_s n}\right)
\end{align}
Finally, we set $\alpha=1$ and $\gamma=1$ and remove the tilde signs in Eqs.(\ref{ds3A},\ref{ds3B}), to obtain
\begin{align} \label{ldot}
    \dot{l}=&v-k(l-1)\left(1-\frac{exp\left(\frac{k(l-1)}{f_s n}\right)}{\kappa n}\right)\\ \label{ndot}
    \dot{n} =& r(1-n)-n\cdot exp\left(\frac{k (l-1)}{f_s n}\right)
\end{align}

Note that the dynamics is now captured by two ODEs, where the spatial component appears only as the total cell length variable. These reduced equations have four structural parameters ($r,k,\kappa,f_s$) and the parameter describing the strength of the actin treadmilling flow ($v$).

\begin{figure*}[htbp!]
\includegraphics[width=\linewidth]{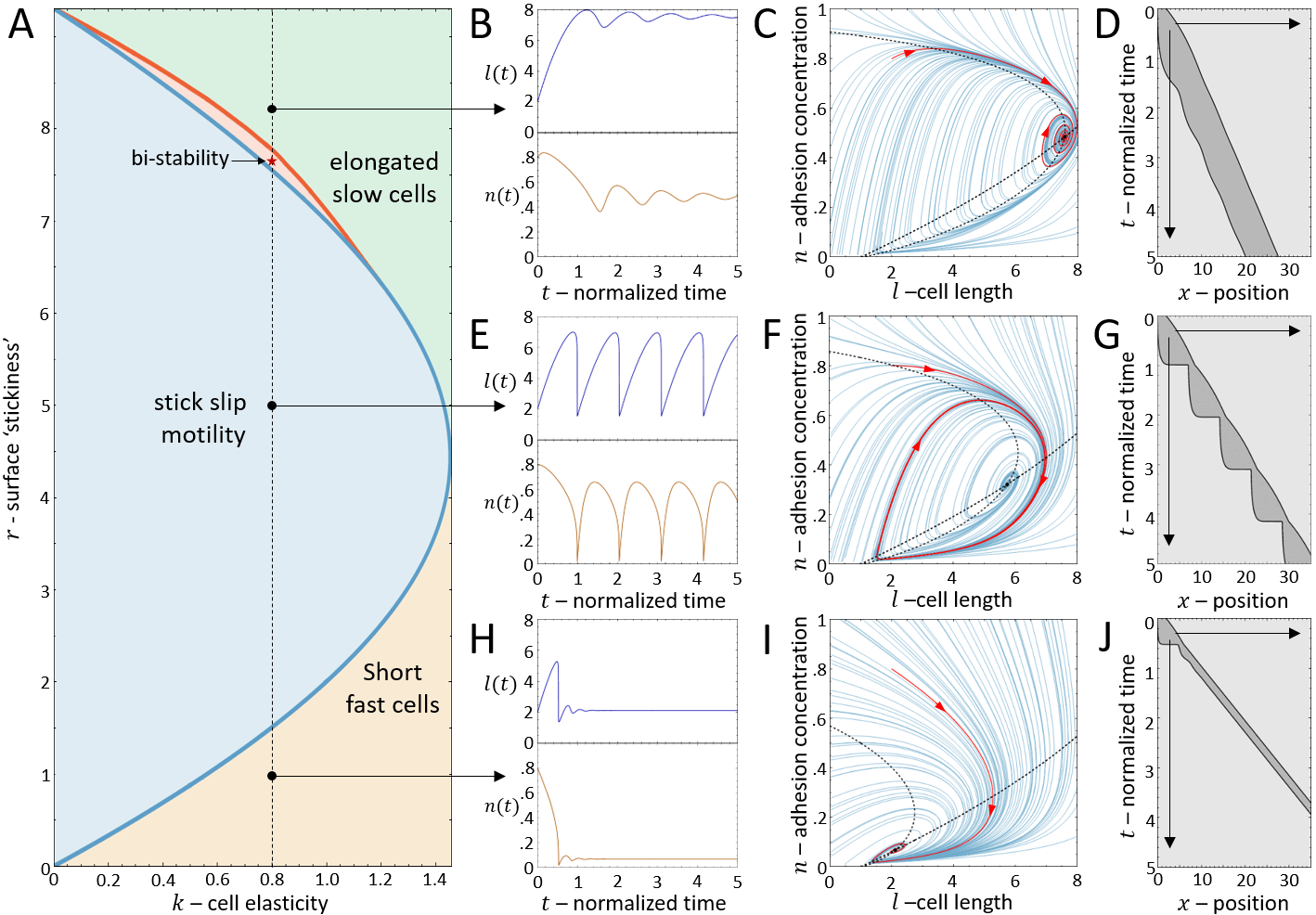}
\caption{\label{fig:constV_02} 
A) $k-r$ phase diagram. Blue region correspond to the polarized stick-slip cells. Orange/Green regions corresponds to the polarized cells with a constant length. Red region corresponds a region of bi-stability between polarized cells with a constant length and stick slip. Blue/Red curves are the hopf/saddle node bifurcation transition lines. Black dashed line is the $r$ cross section for $k=0.8$. Black dots correspond to the points $(k,r)=0.8,1$, $(k,r)=0.8,5$ and $(k,r)=0.8,8.4$. Red star correspond to the point $(k,r)=(0.8,7.7)$. B-D) The dynamics at $(k,r)=0.8,1$. B-D) The dynamics at $(k,r)=0.8,5$. H-I) The dynamics at $(k,r)=0.8,1$. Blue/Orange curves at panels B,E,H correspond to the time series of the cell length and adhesion concentration at the rear. Red curve on panels C,F,I represent the trajectory in $l,n$ phase space. Black curves on panels D,G,I are the kymographs where the black curves are cell edges $x_b$ and $x_f$. Parameters: $f_s=5,\kappa=20,v_f=10$.}
\end{figure*}
\subsubsection*{Results}

It is convenient to plot the resulting dynamics on the $k-r$ phase diagram shown in Fig.\ref{fig:constV_02}A. We find that above a critical value of the cell stiffness $k$ (to the right of the blue solid line), the spring is too stiff to allow for large length changes that enables the spring to store and release large forces. In this regime we get a single, stable fixed point that corresponds to smooth cell motion, and no stick-slip behavior (for the phase-diagram as function of the adhesion saturation parameter $r_0$ see Appendix A).

Below the critical stiffness, we find that for small values of $r$ the system corresponds to smooth motion (stable fixed point), of a short and fast-moving cell (Fig.\ref{fig:constV_02}H-J).

\begin{figure}[htbp!]
\includegraphics[width=\linewidth]{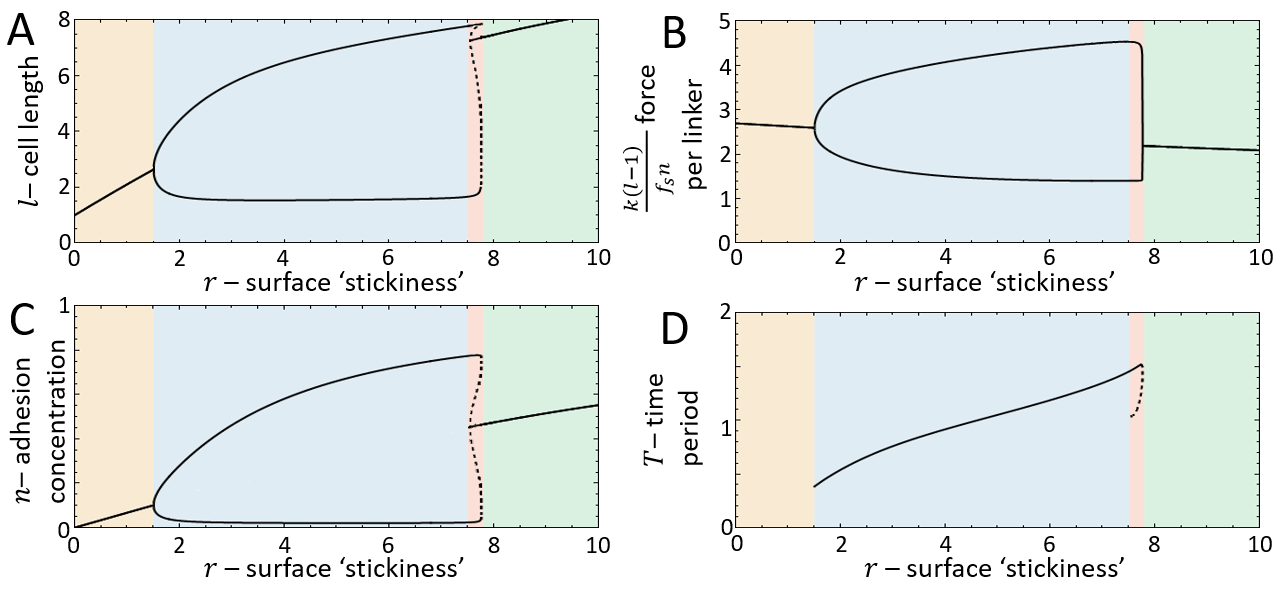}
\caption{\label{fig:constV_04}Stability analysis along the line $k=0.8$ (vertical black dashed line in Fig.\ref{fig:constV_02}A). A) The maximal and minimal length as a function of $r$. B) The maximal and minimal force applied on a linker at the back part of the cell as a function of $r$. C) The maximal and minimal adhesion concentration at the rear as a function of $r$. D) The time period of the limit cycle as a function of $r$. Solid Black curves indicate the stable limit cycle. Dashed black curves indicate the unstable limit cycle.}
\end{figure}

Above a transition line (solid blue line in Fig.\ref{fig:constV_02}A), the fixed point undergoes a Hopf bifurcation, which marks the transition from smooth motion to stick-slip motion (limit cycle dynamics)(Fig.\ref{fig:constV_02}E-G). In this regime there are large length oscillations, as well as large oscillations in the number of bound adhesion linkers at the cell back (Fig.\ref{fig:constV_04}A,B). These oscillations in $n$ occur due to the large values of the force per linker that is reached (Fig.\ref{fig:constV_04}C), which leads to catastrophic, avalanche-like detachment events at the back. Within the stick-slip regime, we find that the duration of the limit-cycle increases with increasing $r$, i.e. the dynamics slow down with increasing substrate adhesiveness (Fig.\ref{fig:constV_04}D).

\begin{figure}[htbp!]
\includegraphics[width=\linewidth]{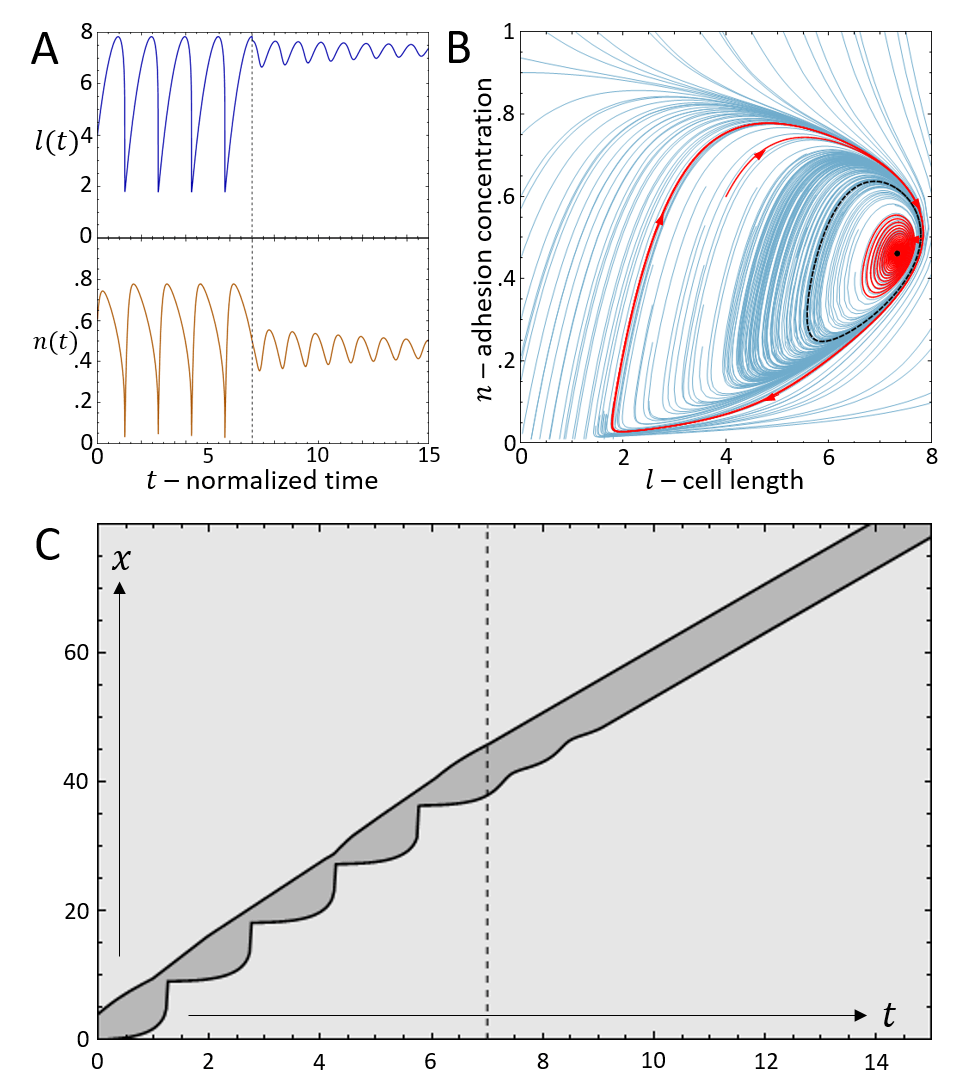}
\caption{\label{fig:constV_03} (A-C) Dynamics in the bi-stability regime (red star in Fig.\ref{fig:constV_02}A) . A) Blue and orange curves are the time series of the cell length and adhesion concentration at the rear. Black dashed line indicates the time when noise was injected into the solution of the differential equation (Eq.\ref{ldot}). B) $n-l$ phase diagram. Red curve is the trajectory and Black curve is the separatrix between the two modes of motion. C) The corresponding kymograph: Black curves represent the edges of the cell $x_b$ and $x_f$. Dashed black line indicates the time when noise was injected.}
\end{figure}
As $r$ increases there is a second Hopf bifurcation, whereby the fixed point is stable again. However, there is a narrow region of bistability, where the phase-space is separated by a separatrix (solid black line in Fig.\ref{fig:constV_04}B), such that the stable fixed point coexists with the limit cycle. Within this regime, noise can induce transitions from stick-slip (limit cycle) to smooth motion (stable fixed point), as demonstrated in Fig.\ref{fig:constV_03}. 

For increasing $r$ the separatrix grows until it meets the limit cycle, which marks the transition to flows that all lead to the single stable fixed point. The dynamics in this regime correspond to a smooth motion of a slow-moving and elongated cell (Fig.\ref{fig:constV_02}B-D,Fig.\ref{fig:constV_04}).

\begin{figure}[htbp!]
\includegraphics[width=\linewidth]{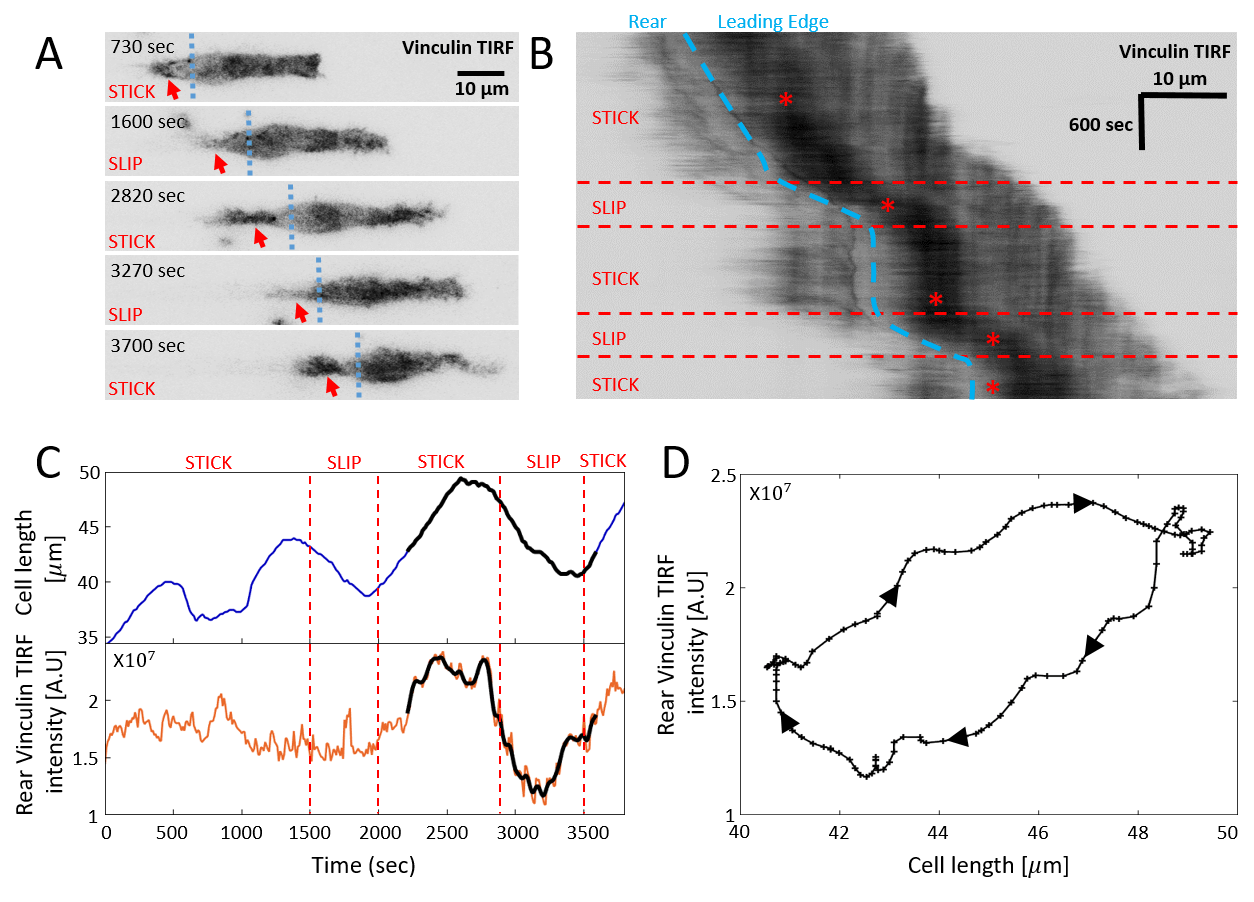}
\caption{\label{fig:constV_06} Comparison of the basic stick-slip model (Eqs.\ref{ldot},\ref{ndot}) to the stick-slip dynamics observed in experiments (Patient-derived Human glioma propagating cell NNI2 transfected with fluorescently taged Vinculin and seeded on laminin-coated lines of 5$\mu$m width and imaged every 110 sec). A) Snap-shots from the movie, where the cells have fluorescently labeled vinculin (represents the extent of adhesion complexes). The region that defines the cell back is to the left of the vertical dashed line, i.e. behind the nucleus. B) A kymograph of the cell migration, with the stick-slip events marked. C) The dynamics of the cell length and the total amount of vinculin signal at the cell back region, as function of time. The black line on both graphs indicates the stick-slip cycle used to plot the phase-space limit-cycle shown in (D).}
\end{figure}

\begin{figure}[htbp!]
\includegraphics[width=\linewidth]{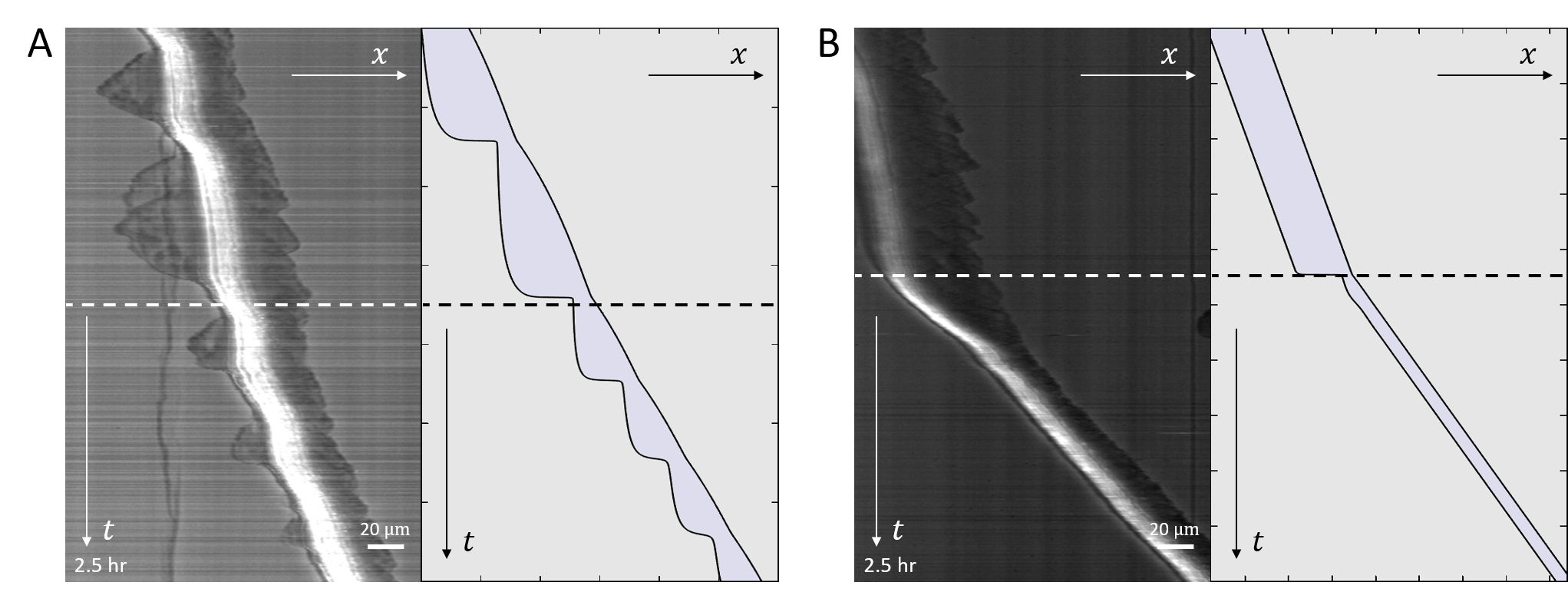}
\caption{\label{fig:constV_05} Comparison of the basic stick-slip model (Eqs.\ref{ldot},\ref{ndot}) to experiments on C6 glioma cells seeded on laminin-coated lines of 5$\mu$m width (imaged every 30 sec). Kymographs correspond to total time 2.5 hours. A) Comparison to model results with $r=7$, then abruptly changed to $r=3$ (indicated by the horizontal dashed line). B) As in (a), with $r=8.5$ then changing abruptly to $r=1$.}
\end{figure}
\subsubsection*{Comparison to Experiments}

We now compare in detail the model's stick-slip limit-cycle dynamics (Fig.\ref{fig:constV_02}E-G) to experiments (see Appendix D for experimental methods). In Fig.\ref{fig:constV_06} we analyze the stick-slip migration mode of a C6 glioma cell (seeded on laminin-coated lines of 5$\mu$m width), by following the cell length and the intensity of membrane-bound vinculin, using total internal reflection fluorescence (TIRF) microscopy. As a component of the adhesion complex, we denote the total intensity of vinculin in the rear of the nucleus as a measure of the adhesion strength at the cell back in the model ($n$). We find that the experimentally observed dynamics during a stick-slip cycle give rise to a limit-cycle in $n,l$-phase space that has the same qualitative features as the model predicts (Fig.\ref{fig:constV_02}F).

The robust features of the migration modes predicted by the model, can be observed in experiments (Fig.\ref{fig:constV_05}). In Fig.\ref{fig:constV_05}a we show a kymograph of a motile cell moving along a one-dimensional stripe, exhibiting stick-slip behavior. At a certain point along the trajectory the cell increased both its rate of stick-slip events, and its overall migration velocity. In addition, the amplitude of the stick-slip events decreased in size. All of these features are captured by the model, if the average adhesion strength between the cell and the substrate ($r$) decreases abruptly along the trajectory, and the cell moves from high to low $r$ within the stick-slip regime (Fig.\ref{fig:constV_02}A). The cell-substrate adhesion strength is sensitive to the surface properties along the stripe, which may vary along its length. The backwards growth of the lamellipodia at the cell back, during each stick-slip cycle, are described by the more elaborate model below.

In Fig.\ref{fig:constV_05}b we show a cell migrating smoothly, and then abruptly decreasing in length and increasing in speed. This behavior is captured by the model assuming that the adhesion strength $r$ decreases such that the cell jumps from high to low $r$ as shown in (Fig.\ref{fig:constV_02}A,D,J).

Note that we can not exclude that the abrupt change in behavior observed for the cells in Fig.\ref{fig:constV_05} originates from a change in some other internal parameter of the cell. In addition, we observe highly dynamic stick-slip behavior at the leading edge of the cell, which we do not describe by our current model \cite{chan2008traction}.

\subsection*{\label{sec:model1}Part 2: Polarized cell with a dynamic protrusion}
\subsubsection*{Model description}
\begin{figure}[h]
\includegraphics[width=\linewidth]{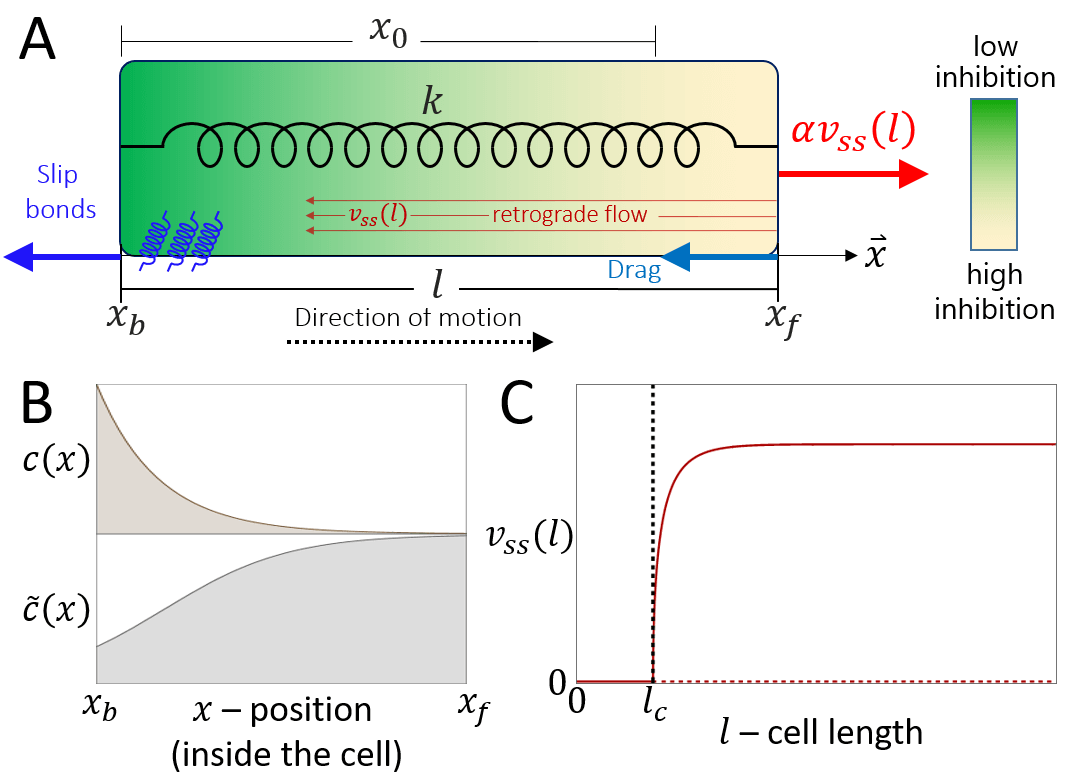}
\caption{\label{fig:dynamicV_01}The stick-slip UCSP model. A) Illustration of a the physical model. Unlike the simple model of Fig.\ref{fig:constV_01}, the treadmilling velocity is now dependent on the cell length  $v(l)$, due to the advection of an inhibitory cue. The concentration of the inhibitory cue is represented by the colorbar on the right. The steady-state concentration of inhibitory cue, $c(x)$, is shown in (B, upper panel), which is an exponential function due to a balance between diffusion and advection (Eq.\ref{cx}). The effect of $c(x)$ on the local actin polymerization rate is given by a Hill function (Eq.\ref{vss}) ((B) lower panel): where the inhibitory cue concentration is low the local actin polymerization rate is large. C) The steady state retrograde flow speed as a function of cell length (solution of Eq. \ref{vss2}). Red solid line represent the stable solution. Red dashed line represent the unstable solution. Black dashed line notes the critical length $l_c$ of polarization.}
\end{figure}

We now complement the length-adhesion model described above, by incorporating a model for the spontaneous self-polarization of the cell. We use a modified version of the scheme developed in \cite{maiuri2015actin} (Fig.\ref{fig:dynamicV_01}A), whereby in addition to the advection of proteins that enhance the actin treadmilling, such as myosin-II, we also consider an inhibitor of actin polymerization. The inhibitor protein is free to diffuse in the cytoplasm, and is also advected by the actin treadmilling flow. An example for an inhibitor of local actin polymerization, that is advected by the actin flow, is Arpin \cite{dang2013inhibitory,maiuri2015actin}. We furthermore assume that the time-scale of the redistribution of this inhibitor across the cell is much shorter than the timescale of changes in the actin treadmilling speed. This assumption is corroborated by recent measurements of myosin-II redistribution time when advected by actin ($\sim10$sec) \cite{tsai2019efficient}, while the stick-slip dynamics of the cell are over timescales of tens of minutes \cite{monzo2016mechanical}.

We can therefore use the steady-state distribution of the concentration of this inhibitor $c(x)$, which is given by an exponential function (Fig.\ref{fig:dynamicV_01}B)
\begin{equation} \label{cx}
c(x)=\frac{c_{tot} v}{D}\left(\frac{e^{-\frac{v x}{D}}}{e^{{-\frac{v x_b}{D}}}-e^{-\frac{v x_f}{D}}}\right)
\end{equation}
where $v$ is the effective instantaneous actin treadmilling velocity, $D$ the effective diffusion coefficient of the inhibitor and $c_{tot}$ is the total amount of inhibitor molecules in the cytoplasm (see Appendix B).

To complete the model \cite{maiuri2015actin}, the net treadmilling flow is given by the difference between the actin polymerization flows created at the two ends of the cell, which are inhibited by the local concentration of the inhibitor at the front and rear 
\begin{align} \label{vss}
v\left(x_f,x_b\right)=&\beta\left(\tilde{c}(x_f)-\tilde{c}(x_b)\right)
\end{align}
where $\beta$ gives the scale of the actin flow in the cell, and the polymerization activity at each end is diminished by the inhibitor concentration at each end, given by a simple Hill function: $\tilde{c}(x_{b,f})=\frac{c_s}{c_s+c(x_{b,f})}$ (Fig.\ref{fig:dynamicV_01}B), where $c_s$ is the concentration at which the effect of the inhibitor saturates.

After rescaling (see Appendix C), using Eqs.\ref{cx},\ref{vss}, we  obtain the following implicit equation for the net actin treadmilling flow
\begin{eqnarray} \label{vss2}
v(l)=\beta\left(\frac{1}{1+c\frac{v}{D}\left(\frac{1}{e^{{\frac{v l}{D}}}-1}\right)}-\frac{1}{1+c\frac{v}{D}\left(\frac{1}{1-e^{-\frac{v l}{D}}}\right)}\right) 
\end{eqnarray}

In the persistent regime (sufficiently large $c_s$), we find that as $l$ increases above a critical value Eq.(\ref{vss2}) undergoes a pitchfork bifurcation, where spontaneous actin treadmilling appears, corresponding to a polarized and motile cell (Fig.\ref{fig:dynamicV_01}C). The critical length is given by
\begin{equation} \label{lc}
l_{c} = \frac{c}{\sqrt{\frac{c\beta}{D}}-1}
\end{equation}
which provides a lower bound for the coupling strength $\beta>D/c$, below which there is no spontaneous motility. A lower critical length for polarization was also obtained in \cite{putelat2018mechanical}.

The equations of motion are now extended beyond the basic model (Eqs.\ref{ldot},\ref{ndot}) to include the dynamics of the actin treadmilling flow
\begin{align}
    \dot{l}=& v-k(l-1)\left(1+\frac{exp\left(\frac{k(l-1)}{n f_s}\right)}{n}\right)\label{ldota}\\
    \dot{n} =& r(1-n)-n\cdot exp\left(\frac{k(l-1)}{n f_s}\right)\label{ndota}\\
    \dot{v} =& -\delta \left(v-v_{ss}(l)\right) \label{vdot}
\end{align}
where $v_{ss}(l)$ is the steady-state solution to Eq.\ref{vss2}, which is the length-dependent actin flow speed, and $\delta$ is the rate at which the actin flow adjusts to length changes, i.e. the rate at which $v$ approaches $v_{ss}(l)$.

For a cell that has a rest-length that is longer than the critical length for polarization, $l_c<1$, the resting state of the cell is polarized. Since the velocity saturates to a constant value for $l>l_c$ (Fig.\ref{fig:dynamicV_01}C), the results are qualitatively similar to those we obtained for a constant, length-independent, $v$ (Fig.\ref{fig:constV_04}).

We will therefore focus on the more interesting case where $l_c>1$, and the rest-length of the cell is smaller than the critical length for polarization. In this case we find another fixed point that corresponds to $v=0$
\begin{equation}\label{v0fixedpoint}
\left(l^*,n^*\right)=(1,\frac{r}{1+r})
\end{equation}
where the cell is at its rest length, and stationary.

\subsubsection*{Results}

\begin{figure*}[htbp!]
\includegraphics[width=\linewidth]{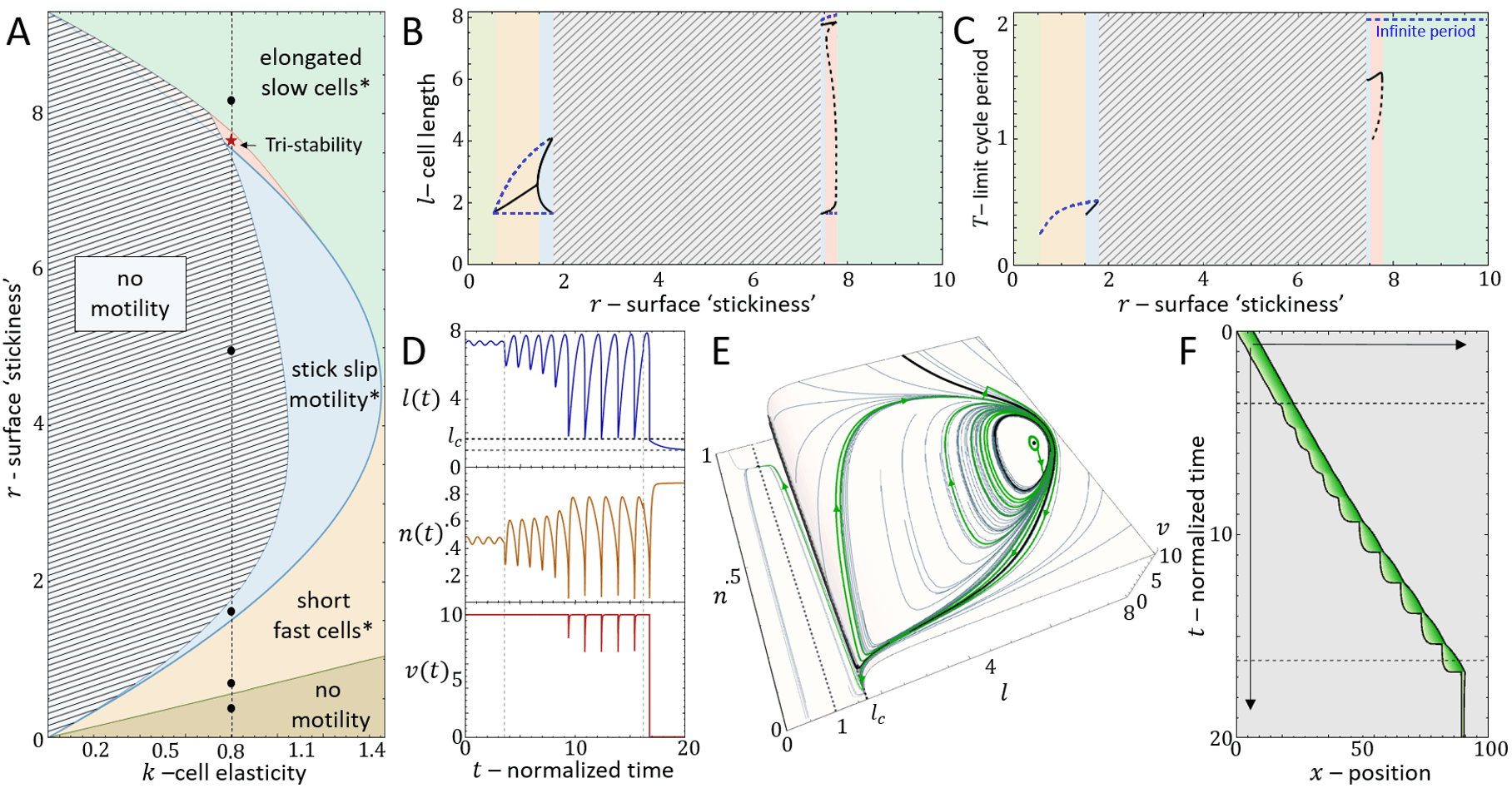}
\caption{\label{fig:dynamicV_02} A) $k-r$ phase diagram. Brown region correspond to a regime where the cell is stationary. Striped texture region corresponds to a regime where there is no motility due to the unstable limit cycle created by the discontinuous vector field. Blue region correspond to a bi-stable region of polarized stick-slip cells and stationary cells. Orange/Green regions correspond to bi-stable regions  of polarized cells with a constant length and stationary cells. Red region corresponds to a region of tri-stability between polarized cells with a constant length, stick slip motility and no motility. Blue/Red/Brown curves are the hopf/saddle node/hopf bifurcation transition lines. Black dashed line is the $r$ cross section for $k=0.8$. Black dots correspond tp the points $(k,r)={(0.8,0.4),(0.8,0.7),(0.8,1.65),(0.8,5),(0.8,8)}$. Red star correspond to the point $(k,r)=(0.8,7.7)$. B) Maximal and minimal amplitude of the $l$ state variable as a function of $r$ for the limit cycles in the vector field along the cross section of $k=0.8$. Dashed blue curve correspond to the unstable discontinuous limit cycle. Solid Black curve corresponds to the stable limit cycle and the cell length. Black dashed curve is the unstable limit cycle. C) The time period of the limit cycles. Blue dashed line indicate the discontinuous unstable limit cycle. Black solid line indicate the stable limit cycle (period of stick slip). Black dashed line indicates the unstable limit cycle in the tri-stable region.
D-F) The tri-stable region. Solution of Eqs.\ref{ldota},\ref{ndota} with noise of amplitude $(\delta l,\delta v)=(0.5,0.11)$ injected at $t=3.55$ and $(\delta l,\delta n)=(0.5,0.05)$ injected at $t=16.2$, to demonstrate the transition between the phases (across the separatrix lines). Blue/Orange/Red curves in D correspond to the dynamics of length/adhesion concentration and actin retrograde flow. Gray dashed lines indicate the time point in which the noise was injected. Green curve in E is the trajectory in the $l-n-v$ phase space. Black solid lines are the separatrices. F displays the kymograph which corresponds to D-E. Parameters: $f_s=5$,$\kappa=20$,$\beta=11$, $c=3.85$, $d=3.85$.}
\end{figure*}

In Fig.\ref{fig:dynamicV_02}A we plot the $k-r$ phase diagram for the simpler case of $\delta\rightarrow\infty$, such that the actin flow speed is given directly by the solution $v_{ss}(l)$ (Eq.\ref{vss}, Fig.\ref{fig:dynamicV_01}C). In comparison with the case of fixed actin flow (Fig.\ref{fig:constV_02}A), there are several new phases, despite the overall similar shape of the phase diagram.

\begin{figure*}[h]
\includegraphics[width=\linewidth]{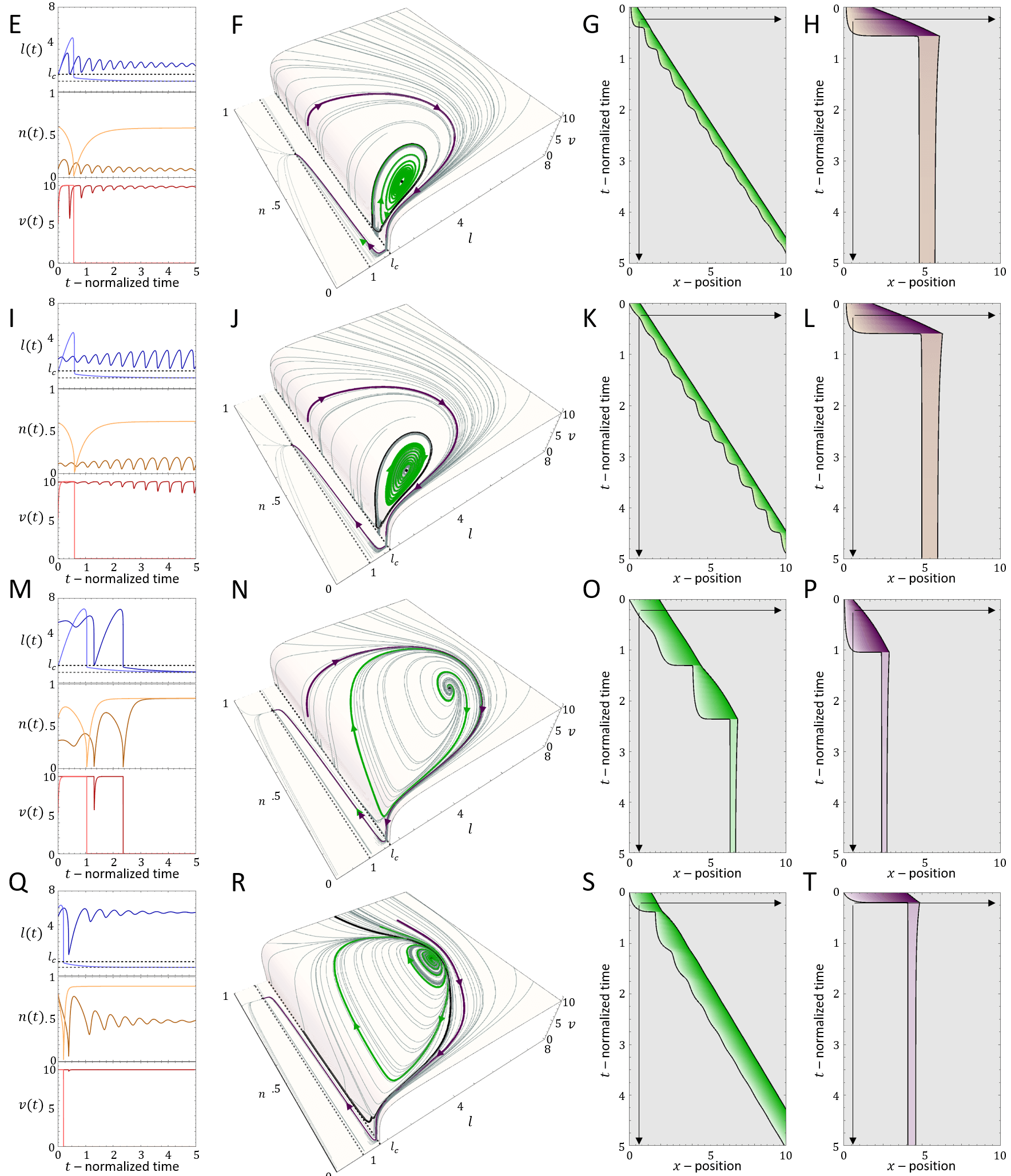}
\caption{\label{fig:dynamicV_03}The dynamics along the line of constant $k=0.8$ (vertical dashed line in Fig.\ref{fig:dynamicV_02}A). A-D) $r=0.4$. E-H) $r=0.7$. I-L) $r=1.65$. M-P) $r=5$. Q-T) $r=8$. Blue/Orange/Red curves in panels A,E,I,M,Q correspond to the time series of the cell length, adhesion concentration and actin retrograde flow respectively (bold/thin lines corresponds to green/purple trajectory). Green and purple lines on panels B,F,J,N,R demonstrate the trajectories in the bi-stable $l-n-v$ phase space regime. Black solid curves are the separatrices. Panels D,G,I display the corresponding kymographs. Parameters: $f_s=5,\kappa=20,\beta=11$,$c=3.85$,$d=3.85$.}
\end{figure*}

For very low values of $r$, we find that there is a transition line below which the cell is non-motile (Brown region in Fig.\ref{fig:dynamicV_02}A), which corresponds to all the flows in the $n-l$ phase-space leading to the single fixed point at $v=0$ (Eq.\ref{v0fixedpoint}). Above this transition line, with increasing $r$, there are phases of coexistence between smooth motion or stick-slip, and non-motile behavior. We plot in Fig.\ref{fig:dynamicV_03} the dynamics for different points of increasing value of $r$, for fixed $k=0.8$ (vertical dashed line in Fig.\ref{fig:dynamicV_02}A). Two flows are demonstrated for each case (green and purple paths and corresponding kymographs in Fig.\ref{fig:dynamicV_03}), exposing the coexistence of either smooth motion (E-H, Q-T) or stick-slip (I-L) with the non-motile phase. The bifurcations of the stable and unstable solutions are denoted by their cell length (and the limit cycle amplitude), as well as the limit cycle periods, in Fig.\ref{fig:dynamicV_02}B,C (along the same $k=0.8$ line).

There is even a thin region of tri-stability (Fig.\ref{fig:dynamicV_02}A), which we demonstrate in Fig.\ref{fig:dynamicV_02}D-F, by inducing a transition between the three phases by adding noise at specific times. For a fuller exploration of the $k-r$ phase diagram see Appendix E.

Next, we explore the effect of a finite value of $\delta$, where the treadmilling velocity of the actin does not instantaneously adjust to its length-dependent value $v_ss(l)$ (Eq.\ref{vdot}). 

\begin{figure*}[htbp!]
\includegraphics[width=\linewidth]{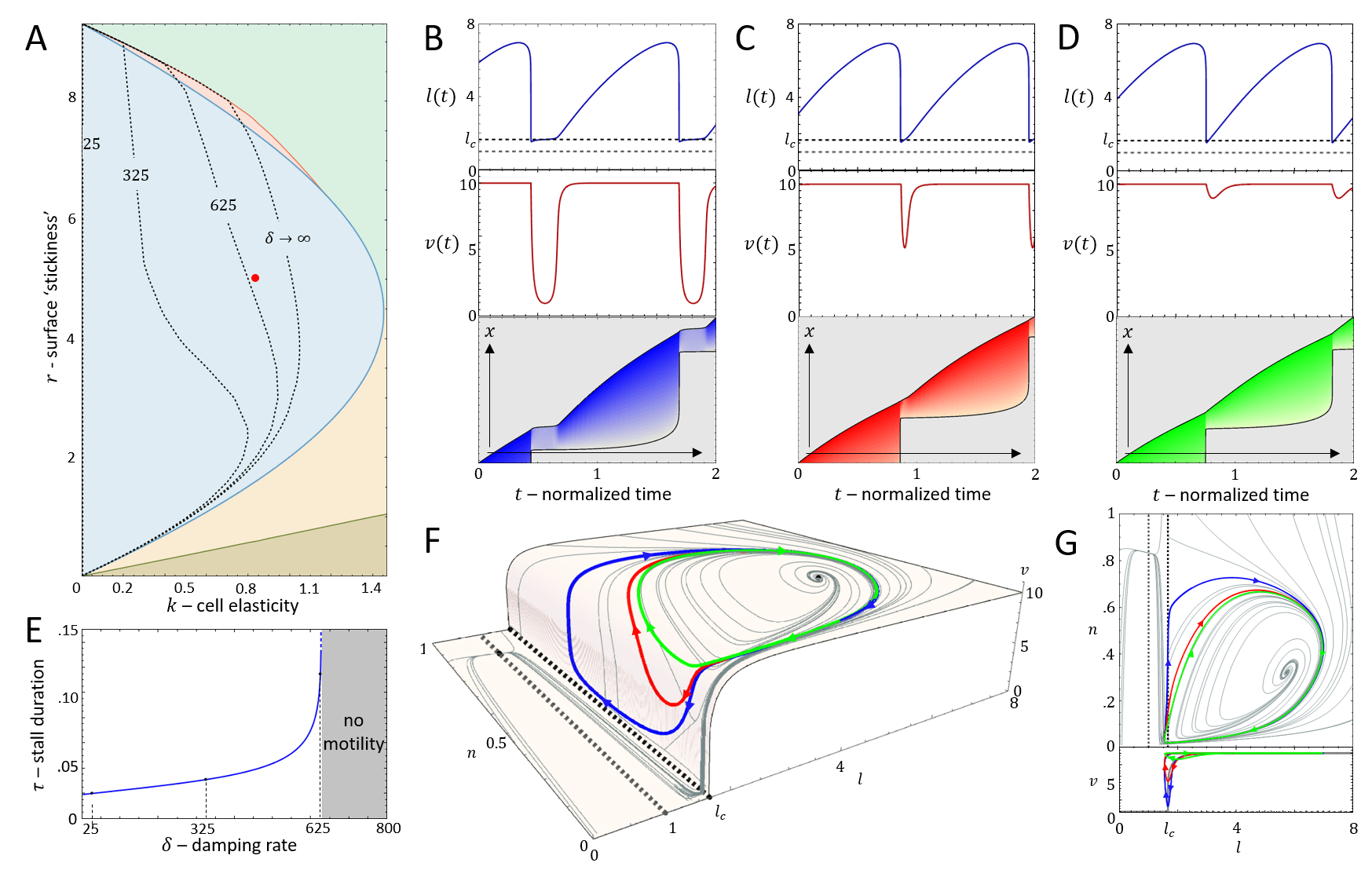}
\caption{\label{fig:dynamicV_04}The effect of a finite actin response rate $\delta$ on the dynamics (Eq.\ref{vdot}). A) The $k-r$ phase diagram for different values of $\delta$ (in Eq.\ref{vdot}). Dashed black lines indicate the coalescence of the discontinuous unstable and stable limit cycles, which produce the no-motility regime for different values of $\delta$. The phase transition line for $\delta\rightarrow\infty$ is the same as in Fig.\ref{fig:dynamicV_02}A. B-D) Demonstration of the lagging time for values of $\delta$ at the point $(k,r)=(0.8,5)$ on the phase diagram (black dot). Blue and red curves are the time series of the cell length and actin retrograde flow speed, respectively. Black dashed line is the critical length of polarization and gray dashed line is the rest length of the cell. Blue, red and green kymographs correspond to $\delta=625, 325$ and $\delta=25$ respectively. E) The duration of the stall as a function of $\delta$,  indicating the values of $\delta=625,325,25$ at the point $(k,r)=(0.8,5)$. Gray region indicates where there is no motility for every initial condition. F) The stable limit cycles which correspond to panels B-D in the $l-n-v$ phase space. G) Projections of the stable limit cycles which correspond to panels B-D in the $l-n$  and $l-v$ phase spaces. Other parameters: $f_s=5,\kappa=20,\beta=11$,$c=3.85$,$d=3.85$.}
\end{figure*}

In Fig.\ref{fig:dynamicV_04}A we plot the shift in the stick-slip transition line to lower values of $k$, for decreasing values of $\delta$. The region of no-motility is now pushed to lower values of $k$. The reason for this is shown in Fig.\ref{fig:dynamicV_04}B-D: when the length drops below $l_c$ after the slip event, the velocity of the actin does not drop to zero instantaneously (as in the case when $\delta\rightarrow\infty$). There is therefore a region of the phase-diagram where the length recovers and increases above $l_c$, allowing the stick-slip limit-cycle to survive. The "stall duration", during which $l<l_c$ and the cell is almost stalled, increases with increasing value of $\delta$ (Fig.\ref{fig:dynamicV_04}E).


\subsection*{\label{sec:model2}Part 3: Self Polarized symmetric cell with a dynamic protrusion}
\subsubsection*{Model description}
In the model we developed so far, we did not allow for the actin polymerization to produce local traction and protrusive forces at both ends, but rather assumed that the competition between the two ends gives rise to a single leading edge (or to an unpolarized cell) (Fig.\ref{fig:dynamicV_01}A).

\begin{figure}[h]
\includegraphics[width=\linewidth]{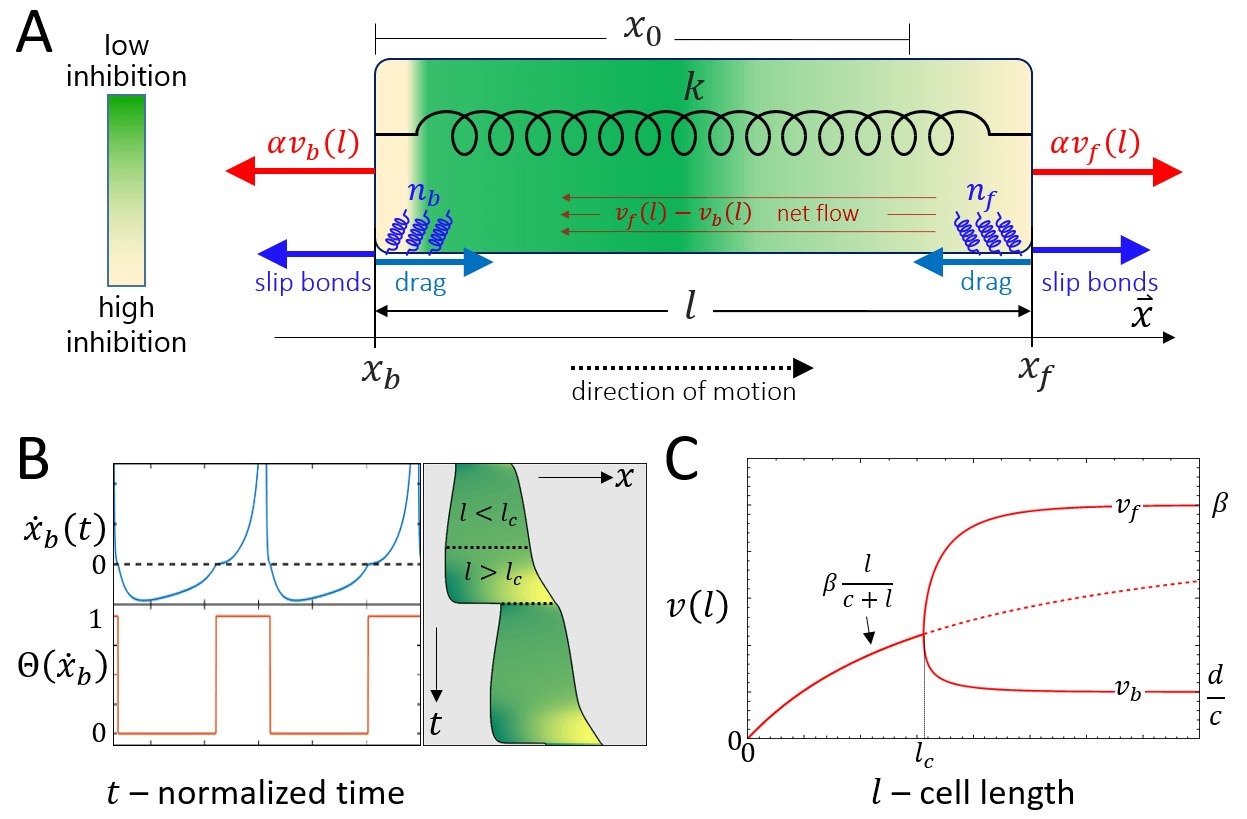}
\caption{\label{fig:symV_01}The symmetric model. A) Illustration of the physical model.  $x_b$ and $x_f$ are the back and front part of the cell of the cell which are connected by a spring with a stiffness $k$. $x_0$ is the rest length of the spring and $l$ is the length of the cell. $v_f$ and $v_b$ are  the steady state velocities of actin retrograde flow from both ends of the cell, which are coupled to the gradient of an inhibitory cue. The concentration of the inhibitory cue is represented by the colorbar on the left. At both ends of the cell acts a protrusion force (red arrow) which is proportional to the retrograde flow speed (red arrow), a drag force (teal arrow) when the edge is moving forward and a friction force due to the slip bonds (blue arrow) when the edge is moving backwards. B) Demonstration of the friction model in Eq.\ref{gammasym}:  Upper panel displays in blue the derivative of the moving rear $\dot{x}_b$. Below in orange is the Heaviside theta function of the time series of $\dot{x}_b$. Right panel displays the corresponding kymograph. C) The steady state retrograde flow speed as a function of cell length (Eq.\ref{vfvb}). Red solid line represent the stable solution. Red dashed line represent the unstable solution. Black dashed line notes the critical length of polarization $l_c$. Parameters:  $\beta=19$,$c=9$,$d=40$.}
\end{figure}

We now extend the model to describe the local protrusive forces that are produced by actin polymerization at the two ends of the cell (Fig.\ref{fig:symV_01}A).

The dynamics of the moving front and back depend on their respective direction of motion: when they protrude there is a drag force that is given by a constant drag coefficient, while when they retract the friction depend on the slip-bond activity of the adhesion (Fig.\ref{fig:symV_01}B). We therefore have the following equations of motion for the two ends of the cell
\begin{equation}\label{xfb}
    \dot{x}_{f/b}= \frac{1}{\Gamma_{f/b}}\left(\pm v_{f/b}\mp k(x_f-x_b-1)\right)
\end{equation}
where the friction coefficients are given by
\begin{eqnarray}
    \Gamma_{f/b}=&1-\Theta\left(\mp\dot{x}_{f/b}\right) \nonumber\\
    +&\Theta\left(\mp\dot{x}_{f/b}\right)n_{f/b} \kappa\cdot exp\left(\frac{-v_{f/b}+k(x_f-x_b-1)}{f_s n_{f/b}}\right) \label{gammasym}
\end{eqnarray}
where $\Theta$ is the theta Heaviside function. This way the friction coefficient depends on the local motion of each end. This is illustrated in Fig.\ref{fig:symV_01}B, where the friction at the back is due to the slip-bonds when the back end of the cell is retracting $\dot{x}_b>0$ and $\Theta\left(\dot{x}_{b}\right)=1$.

Note that the force that pulls on the adhesions during retraction (in the exponential function in Eq.\ref{gammasym}) is not only given by the elastic spring force (as in the previous models presented above), but is countered by the local force of actin polymerization that is pushing the membrane

Similarly, the dynamics of the adhesions are given by
\begin{equation}
    \dot{n}_{f/b}=r(1-n_{f/b})- n_{f/b}\cdot exp\left(\frac{-v_{f/b}+k\left(x_f-x_b-1\right)}{f_s n_{f/b}}\right)
\end{equation}
The dynamics of the polymerization/tread-milling of the actin is now calculated separately on both sides of the cell, given by
\begin{equation}
    \dot{v}_{f/b}=-\delta\left(v_{f/b}-v^{*}_{f/b}\right) \label{vbfdelta}
\end{equation}
where 
\begin{equation} \label{vfvb}
    v_{f/b}^{*}=\beta\frac{c_s}{c_s+\tilde{c}\left(x_{f/b}\right)}
\end{equation}
The distribution of the inhibitor along the cell is affected by the competition between the treadmilling actin from both ends, so its given by Eq.\ref{cx} with: $v=v_f-v_b$. 

\subsubsection*{Results}

The solution of Eq.\ref{vfvb} is shown in Fig.\ref{fig:symV_01}C. The solution shows that the cell remains unpolarized for $l<l_c$, and the actin tredmilling flows are equal at the two ends of the cell, given by $v_{f/b}\beta\left(\frac{l}{c+l}\right)$. A polarized cell for $l>l_c$ has a higher treadmilling flow from the front, and in the limit of large $l$ the treadmilling velocity approaches these limiting values: $v_f^*=\beta$ and $v_b^*=d/c$.

\begin{figure*}[h]
\includegraphics[width=\linewidth]{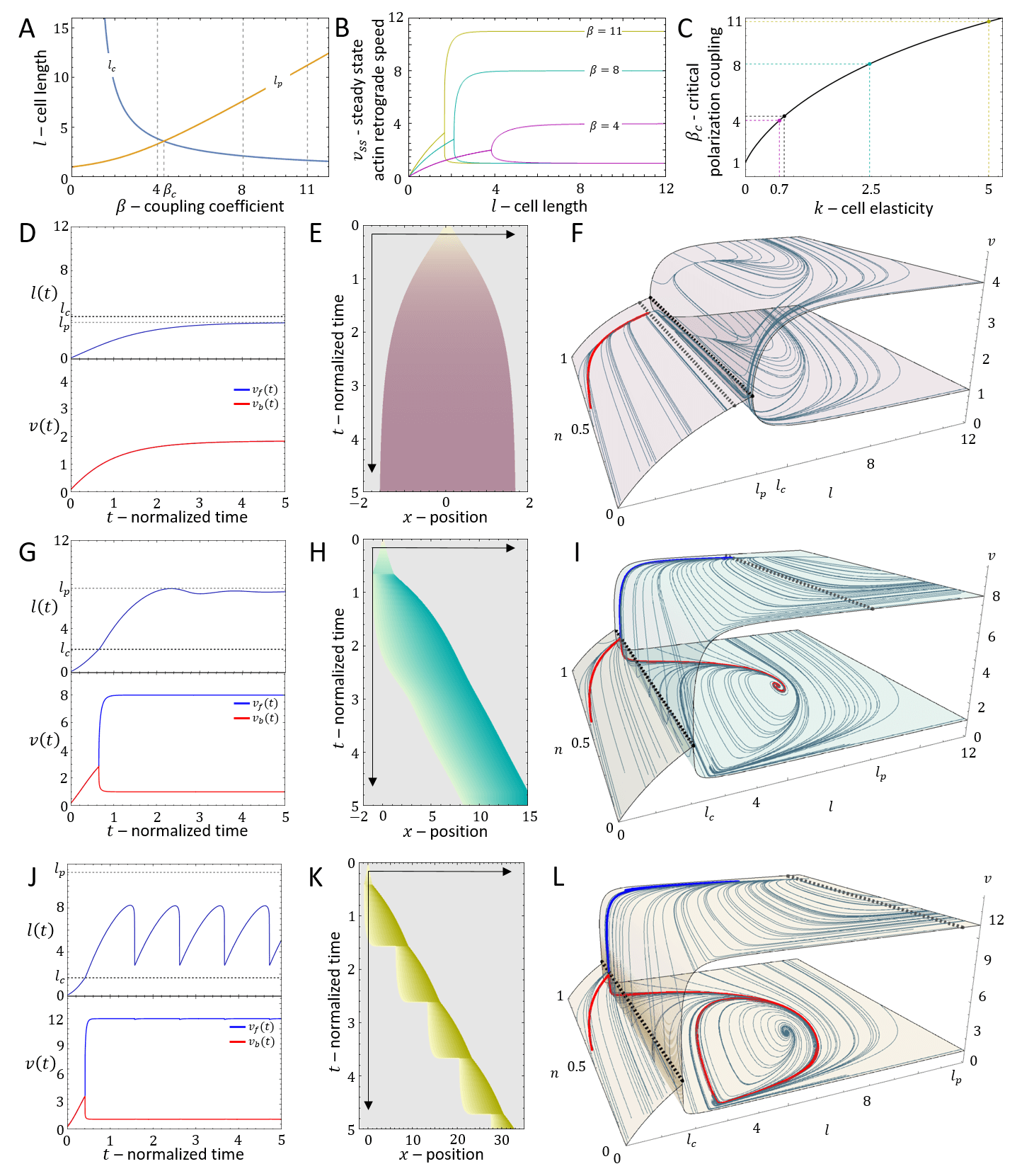}
\caption{The polarization length. A) The critical length $l_c$ (blue) and polarization length $l_p$ (orange) as a function of the coupling coefficient $\beta$. Gray dashed lines indicate the sample points of $\beta=4,8,11$
. B) The steady state velocity profile as a function of length for $\beta=4,8,11$ (purple, teal, yellow curves). C) The critical coupling coefficient as a function of cell elasticity $k$. The purple, teal, yellow dashed curves denote the critical values of $k$ for the values of $\beta$ used in (B,D-L). The black dashed line denotes the critical value of $\beta$ for $k=0.8$ (which is the value of $k$ used in D-L). D-F) Time series, kymograph, and phase space for $\beta=4$. G-I) Time series, kymograph, and phase space for $\beta=8$. J-L) Time series, kymograph, and phase space for $\beta=11$. In panels D,G,I upper blue curve is the time series of the cell length. Dashed black/gray lines are the critical/polarization lengths. Lower red/blue curves represent the time series of the actin retrograde flow speed at the front/back respectively. In panels E,H,K the purple/teal/yellow colors represent $\beta=4,8,11$. In panels F,I,L red/blue curves represent the trajectories at the $l-n_f-v_f$/$l-n_b-v_b$ phase spaces (upper/lower part of the surface). Parameters:  $\kappa=20$, $f_s=5$, $r=5$, $k=0.8$, $c=3.85$, $d=3.85$.} \label{fig:symV_02}
\end{figure*}

We find that there a new length scale in the system, which determines the ability of the cell to elongate and reach polarization. For the the unpolarized cell we equate the protrusive forces at both ends to the restoring force of the spring 
\begin{equation}
    k(l-1)=\beta\left(\frac{l}{c+l}\right)
\end{equation}
which yields the polarization length
\begin{equation} \label{lp}
    l_p=\frac{1}{2}(1-c)+\frac{\beta}{2k}+\sqrt{c+\left(\frac{1}{2}(c-1)-\frac{\beta}{2k}\right)^2}
\end{equation}
By equating the critical length $l_c$ (\ref{lc}) to the polarization length $l_p$ (\ref{lp}) we obtain a critical value of $\beta$, above which the cell polarizes (Fig.\ref{fig:symV_02}A,B)
\begin{equation}
    \beta_c = \frac{d}{2c}+ck+\frac{c^2k^2}{2d}+\left(\frac{ck-d}{2c d}\right)\sqrt{d^2+2c(1+2c)d k+c^2k^2}
\end{equation}

As for the previous model, the interesting behavior is for a cell that has a rest-length that is smaller than $l_c>1$. For such a system we demonstrate the dynamics for different values of $\beta$ (above and below $\beta_c$), in Fig.\ref{fig:symV_02}D-L. For $\beta<\beta_c$ the cell elongates symmetrically (Fig.\ref{fig:symV_02}D-F), but does not polarize (no spontaneous symmetry breaking). For $\beta>\beta_c$ we find regimes of smooth motion ($\beta=8$, Fig.\ref{fig:symV_02}G-I) and stick-slip ($\beta=11$, Fig.\ref{fig:symV_02}J-L) for parameters that correspond to these phases in the phase diagram of the stationary treadmilling system (Fig.\ref{fig:constV_02}). 

\begin{figure}[h]
\includegraphics[width=\linewidth]{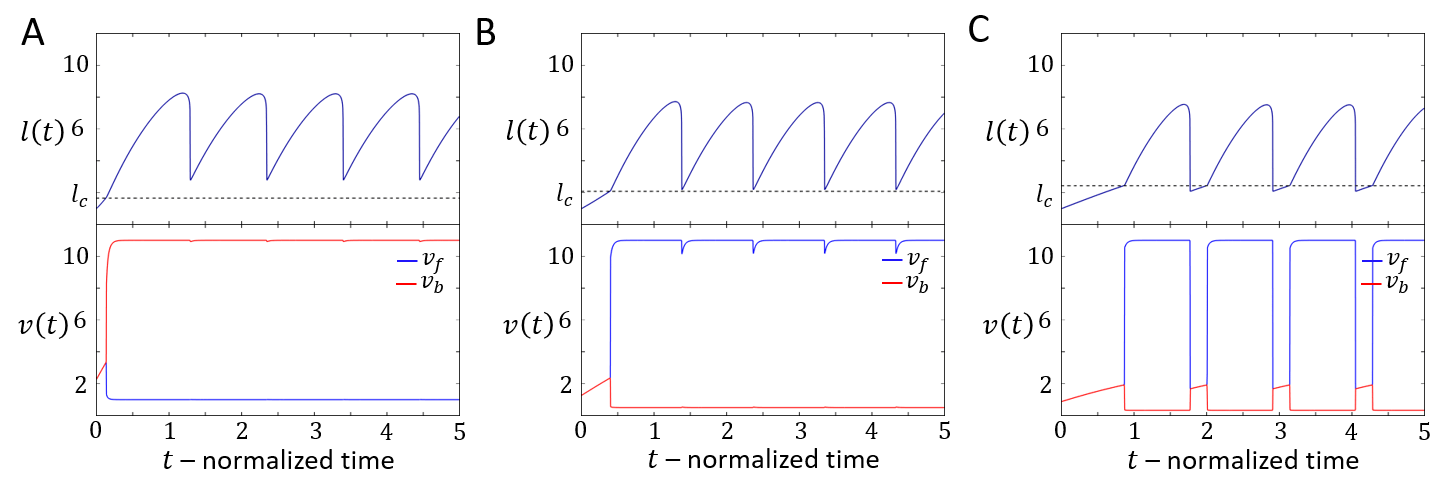} \label{fig:SI_03}
\caption{Time series of the cell length (dark upper panels) and the retrograde flow speed (lower panels) of the symmetric system for different values of the parameter $c$: A) for $c=3.85$. B) for $c=7.7$ and C) for $C=11.55$. Black dashed line in the upper panels is the critical length of polarization $l_c$. Red/Blue curves in the lower panels represent the actin flow at the front/back.}
\end{figure}

When the cell is within the stick-slip regime, it may end up after each slip event with a cell length that is shorter than $l_c$. Since in the symmetric model the protrusive force acts locally at both ends, the cell elongates symmetrically until $l>l_c$ and it repolarizes. This is demonstrated in Fig.\ref{fig:symV_01}B. In Fig.\ref{fig:SI_03} we show the effect of the parameter $c$ on the stick-slip migration, where increasing $c$ decreases the retrograde flow and increases the time it takes the cell to repolarize following each slip event.

\begin{figure*}[h]
\includegraphics[width=\linewidth]{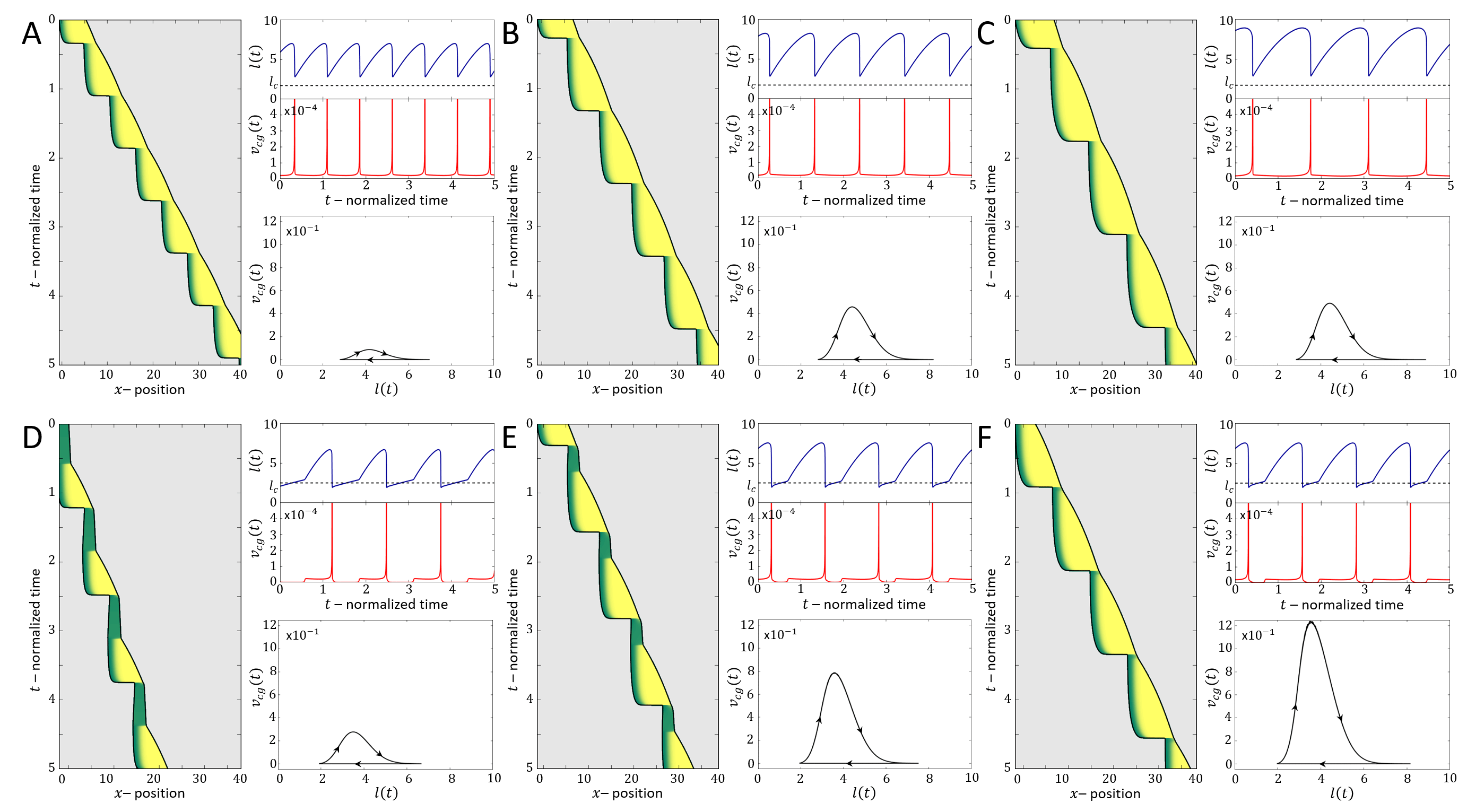} 
\caption{\label{fig:SI_04} The effect of $l_c$ on the overall velocity of migration during stick-slip. A-C) above the critical length. $c=3.95$, $r=3,5,7$. left panel kymograph, right upper panel: cell length and speed time series. lower panel: phase space. Parameters: $\beta = 11$, $d=3.95$, $k=0.8$, $f_s=5$, $\kappa=20$, $\delta=210$. D-E) below the critical length. $c=11.85$, $r=3,5,7$. left panel kymograph, right upper panel: cell length and speed time series. lower panel: phase space. Parameters: $\beta = 11$, $d=3.95$, $k=0.8$, $f_s=5$, $\kappa=20$, $\delta=210$.}
\end{figure*}

\begin{figure*}[h]
\includegraphics[width=\linewidth]{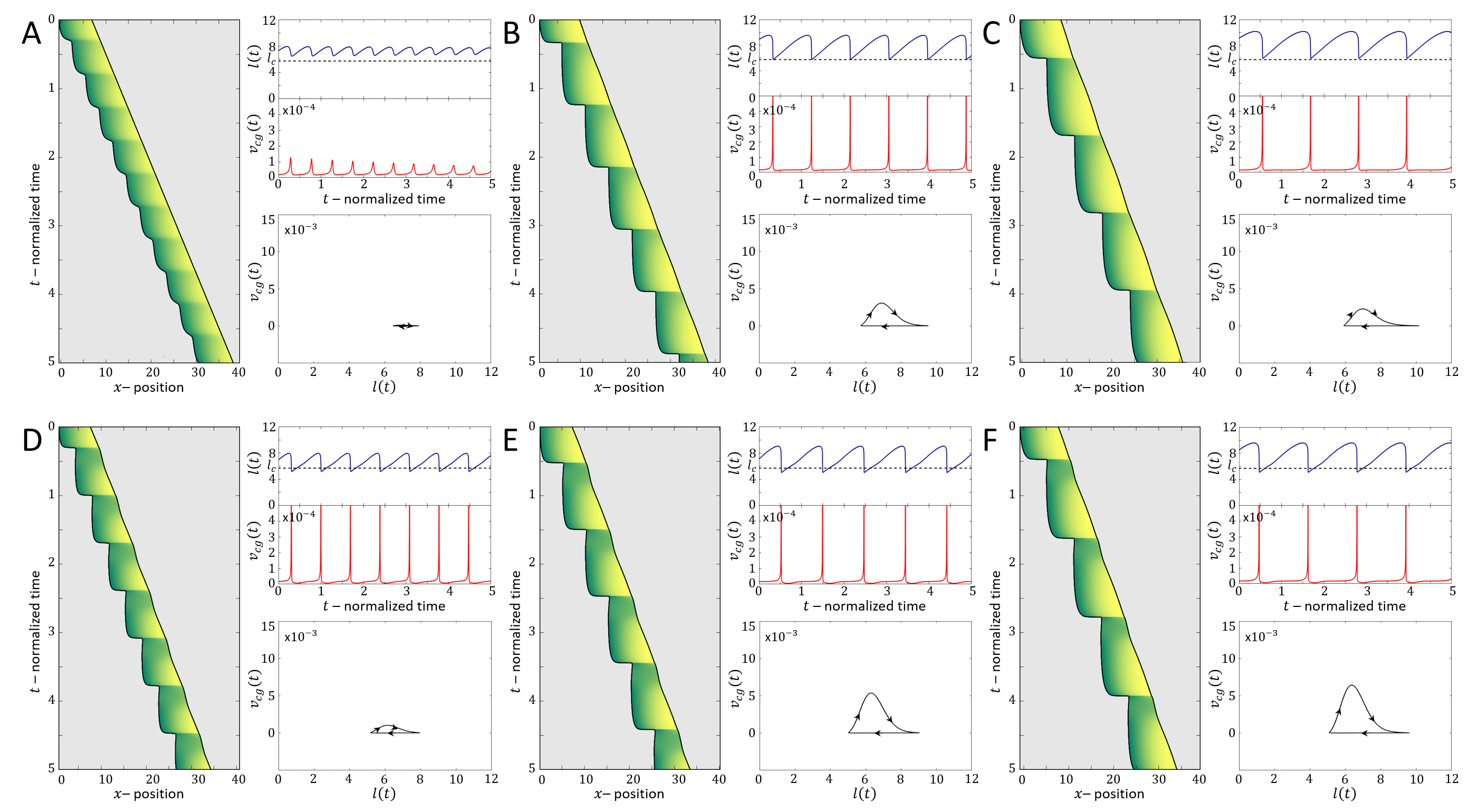} 
\caption{\label{fig:SI_05} Effect of $l_c$ on the overall migration velocity during stick-clip motion. A-C) Cell length is above the critical length. $c=5$, $r=2,4,6$. Left panel: kymograph, right upper panel: cell length and speed time series. Lower panel: phase space. Parameters: $\beta = 11$, $d=3.95$, $k=0.8$, $f_s=5$, $\kappa=20$, $\delta=210$. D-E) below the critical length. $c=6$, $r=2,4,6$. Left panel: kymograph, right upper panel: cell length and speed time series. Lower panel: phase space. Parameters: $\beta = 14$, $d=20$, $k=1$, $f_s=5$, $\kappa=20$, $\delta=120$.}
\end{figure*}

Furthermore, since the cell is effectively stalled when its length is shorter than $l_c$, the time it takes for the cell to expand back to $l_c$ after each stick-slip event determines the overall migration velocity. This is demonstrated in Figs.\ref{fig:SI_04},\ref{fig:SI_05}. 

\begin{figure}[htbp!]
\includegraphics[width=\linewidth]{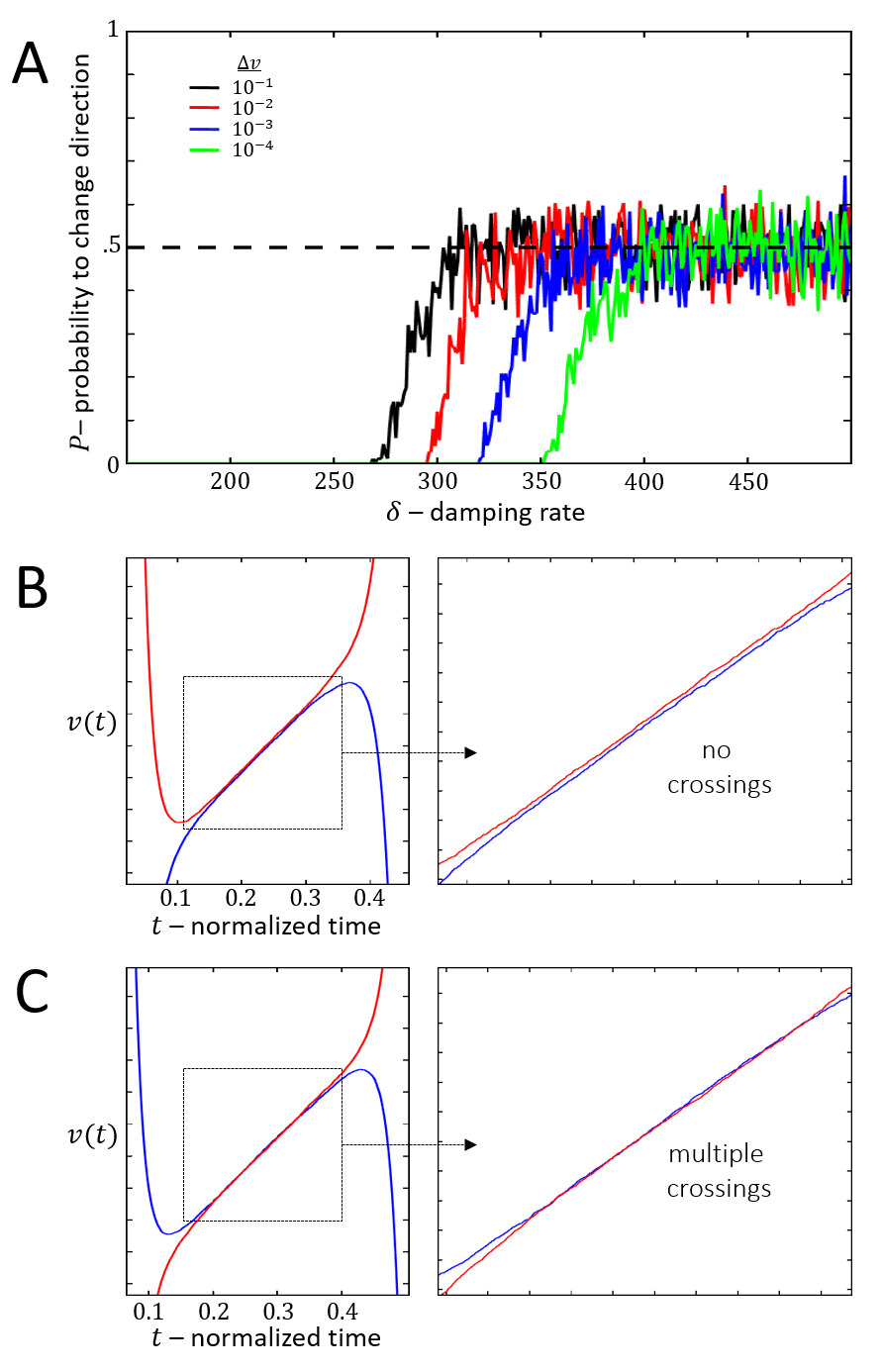}
\caption{A) The probability to change direction as a function of the damping rate for different levels of noise in the actin velocity. B-C) Demonstration of events at which the cell remains polarized/switches direction during the periods where $l<l_c$. Blue/Red curves indicate the actin flow speed at the front/back. Parameters:  $\kappa=20$, $f_s=5$, $r=5$, $k=0.8$, $c=11.55$, $d=3.85$.} \label{fig:symV_04}
\end{figure}

\begin{figure}[htbp!]
\includegraphics[width=\linewidth]{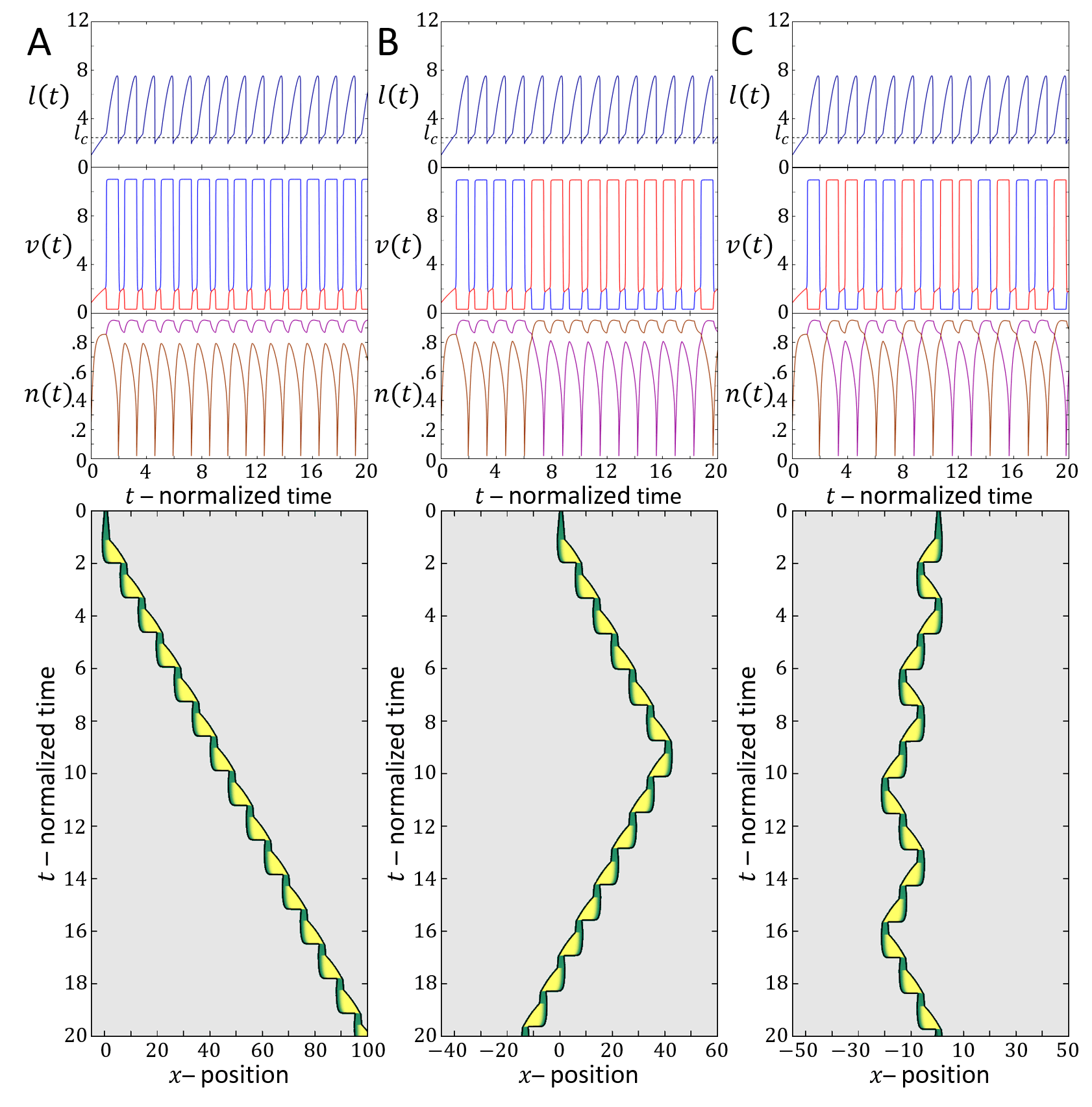}
\caption{Demonstration of direction changes events for different values of $\delta$ with noise amplitude of $\Delta v=10^{-3}$ (see Fig.\ref{fig:symV_03}A). Upper/middle/lower panels correspond to the time series of the cell length/actin flow/adhesion concentration for $\delta=300,325,350$, (A-C) from left to right respectively. At the bottom are the corresponding kymographs. Parameters:  $\kappa=20$, $f_s=5$, $r=5$, $k=0.8$, $c=11.55$, $d=3.85$.} \label{fig:symV_05}
\end{figure}

During each stick-slip event, since the cell loses polarity, it may repolarize in a new direction. This occurs when we add random noise to the equations of motion: the system is solved as a stochastic differential equation where additive noise is added to the actin flow $v_f$ and $v_b$. For different amplitudes of the noise term we plot in Fig. \ref{fig:symV_04}A the probability $p$ that the cell changes its direction of motion per stick-slip event. We find that for each level of noise there is a critical value of the actin response rate $\delta$ (Fig.\ref{vbfdelta}) that sets a threshold between $p=0$ and $p=0.5$. During each stick-slip event, when $l<l_c$, the actin flow velocities on both sides approach each other exponentially with a time-scale of $\delta^{-1}$. As $v_f/b$ approach within the noise amplitude, they cross each other (Fig. \ref{fig:symV_04}B,C), and the side that has the higher flow when $l=l_c$ determines the new direction of cell motion. The dynamics around the critical $\delta$ where the turning probability increases strongly, are demonstrated in Fig.\ref{fig:symV_05}.

\section*{Comparison to experiments}

When comparing the results of our fully symmetric model to the experimental observations of motile cells along one-dimensional tracks, we need to note that the cellular system is very noisy. Nevertheless, our model exposes distinct motility patterns that should underlie the noisy motion of the cells.

\begin{figure}[htbp!]
\includegraphics[width=\linewidth]{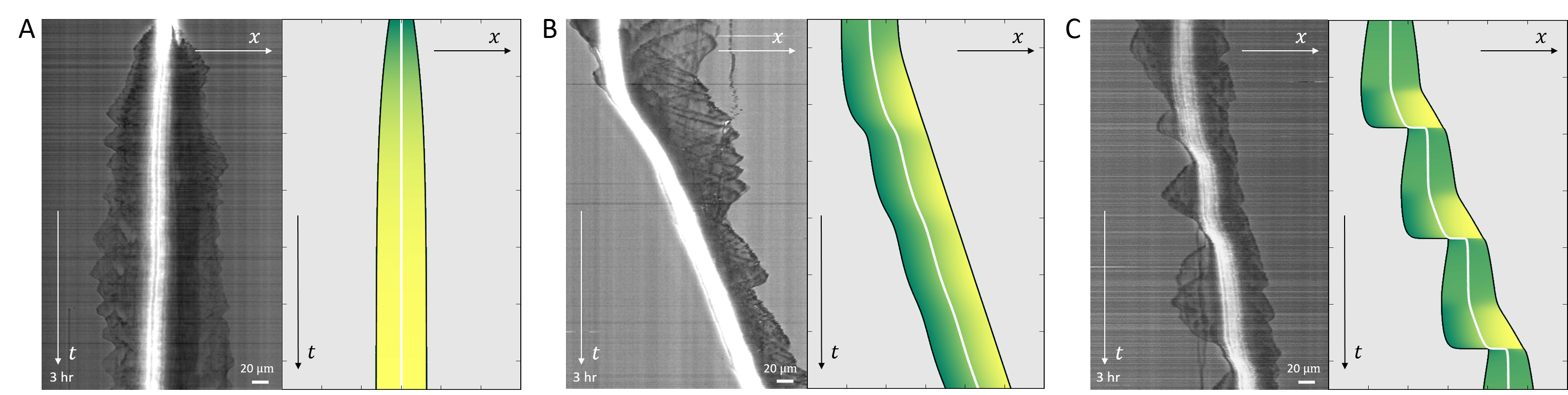}
\caption{\label{fig:symV_03}Demonstrating the different migration modes as a function of the strength of the actin flow ($\beta$), in comparison to observed cell migrations (C6 glioma cells seeded on laminin-coated lines of 5=$\mu$m width, imaged every 30 sec). Kymographs correspond to total time of 3 hours: A) A cell spreading symmetrically and not polarizing: $\beta = 8.5$ (in our model $\beta<\beta_c$, Fig.\ref{fig:symV_02}). B) Smooth migration: $\beta = 9$. Note that the stick-slip at the leading edge observed in the experiments is not included in our model, and does not affect the smooth migration. C) Stick-slip migration: $\beta = 10.5$. Other model parameters:   $\kappa=20$, $f_s=5$, $r=3$, $k=0.75$, $c=7.12$, $d=15.4$. The white line in the model kymographs denotes the trajectory of the cell's geometric center.}
\end{figure}

First, we wish to show that the main classes of cell behaviors, which is observed within a uniform population of identical cells (Fig.\ref{fig:symV_03}) can be qualitatively spanned by the model if, for example, the actin treadmilling activity (model parameter $\beta$) is variable within the population.

\begin{figure}[htbp!]
\includegraphics[width=\linewidth]{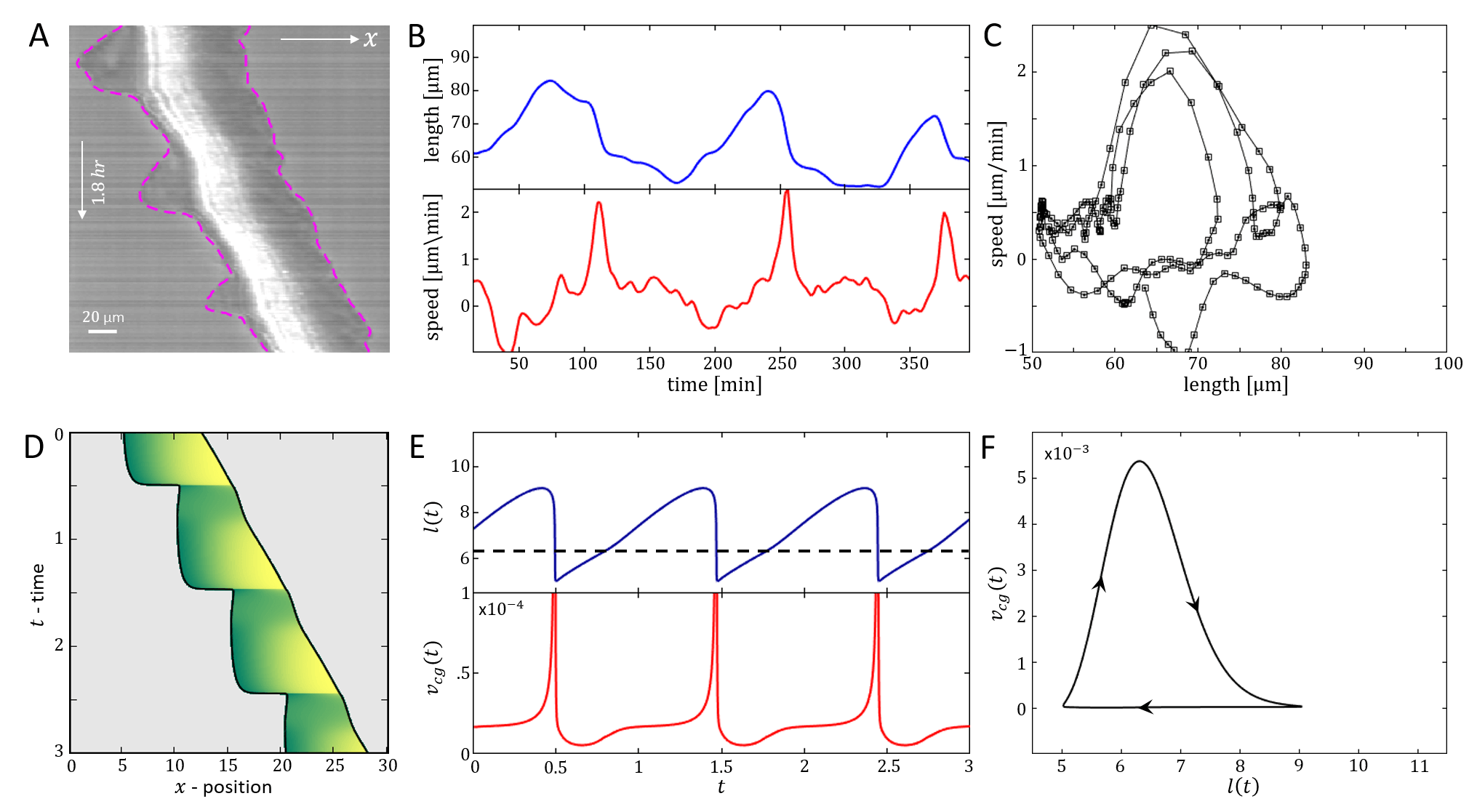}
\caption{\label{fig:expts_01} Stick-slip migration analysis in experiments and the model. A-C) Experiment: C6 glioma cells seeded on laminin-coated lines of 5 $\mu$m width (imaged every 30 sec). Kymograph correspond to total time of 3 hours. A) Kymograph of a cell with stick-slip migration. B) Upper panel is the time-dependent length of the cell. Lower panel is the cell speed (from the trajectory of the geometric center). C) Phase-space trajectory of the cell. D-F) Model results for a cell in the stick-slip regime: A) kymograph. B) length and velocity time series. C) phase space trajectory. parameters: $\beta=14$, $c=6$, $d=20$, $r=4$, $k=0.8$ and $\delta=120$.}
\end{figure}

\begin{figure}[h]
\includegraphics[width=\linewidth]{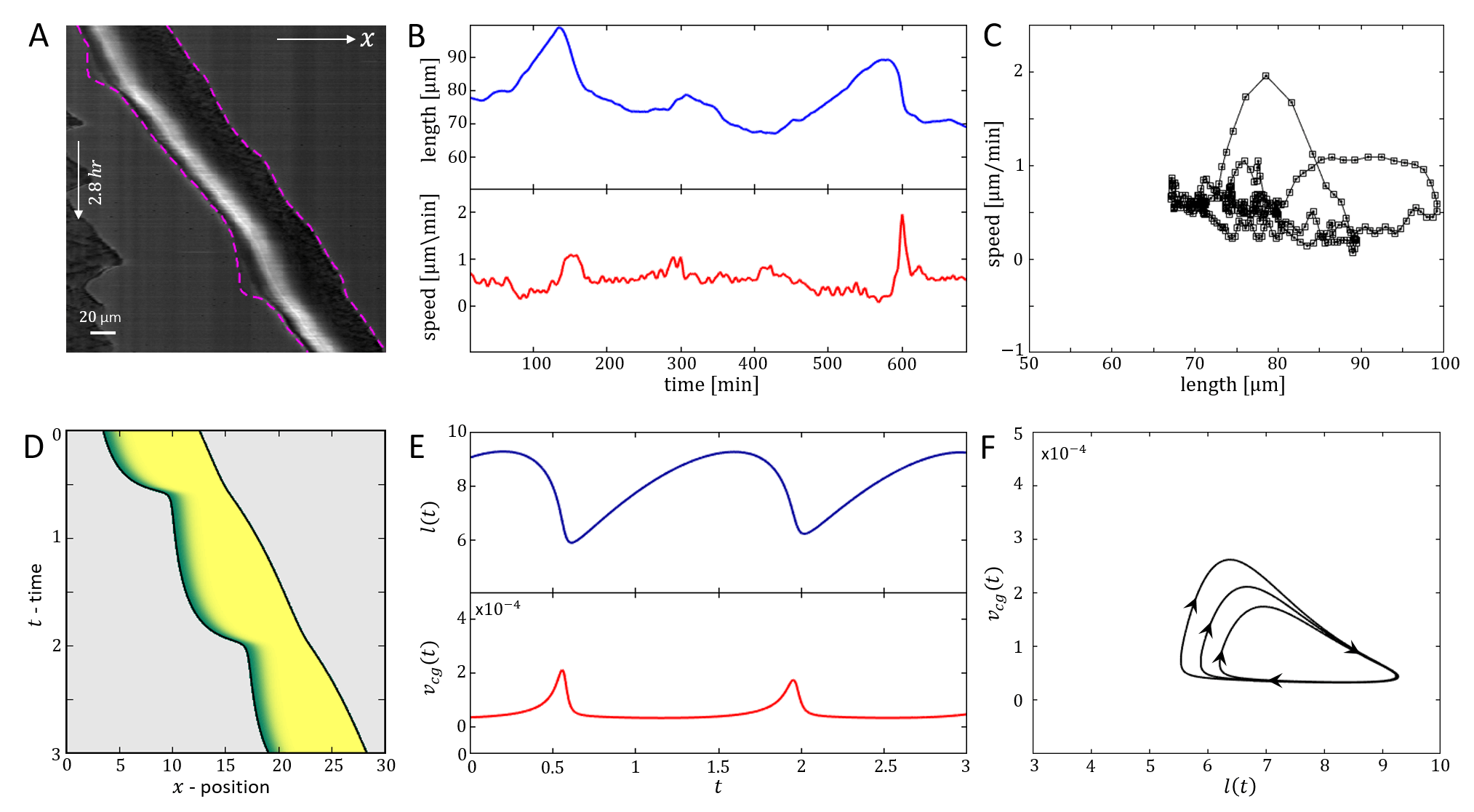} \caption{\label{fig:SI_07}Stick-slip migration comparison to experiment of C6 glioma cells migrating on laminin-coated lines of 5$\mu$m width (frames are 30 sec apart). A-C) Experiment. D-F) Model calculation. For both we plot the kymograph, cell length and speed time series and phase space trajectories. Model parameters: $\beta = 12$, $c=8.37$, $d=10.86$, $k=0.9$, $f_s=5$, $r=6$, $\kappa=16$, $\delta=150$.}
\end{figure}

\begin{figure}[h]
\includegraphics[width=\linewidth]{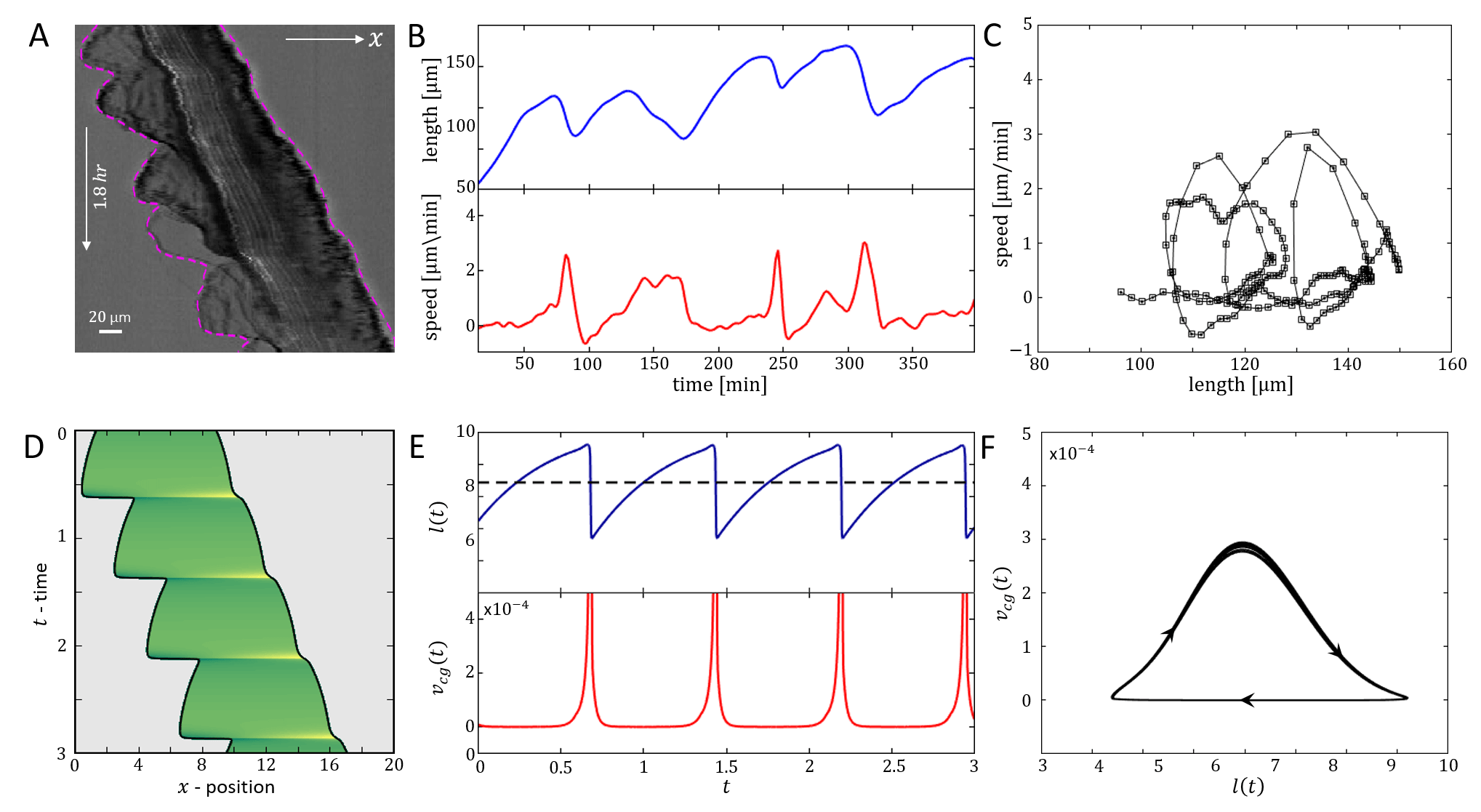} \caption{\label{fig:SI_08}As in Fig.\ref{fig:SI_07}. A-C) Experiment. D-F) Model calculation. Model parameters:  $\beta = 19$, $c=9$, $d=40$, $k=1.1$, $f_s=5$, $r=0.4$, $\kappa=20$, $\delta=120$.}
\end{figure}

Next, we focus on the stick-slip migration pattern, and compare several examples of such observed cell trajectories to the dynamics predicted by the model (Fig.\ref{fig:expts_01}). We find that there is strong similarity between the dynamics exhibited by the cell and in the model, regarding the oscillations in the cell length and velocity. The limit-cycle trajectory predicted by the model can be observed in the experimental data. For slightly different set of parameters, we find that the model gives rise to other stick-slip regimes observed in the experiment: a cell with low-frequency of small-amplitude stick-sip events (Fig.\ref{fig:SI_07}), and a cell with high-frequency of large-amplitude stick-slip events (Fig.\ref{fig:SI_08}).



\begin{figure}[htbp!]
\includegraphics[width=\linewidth]{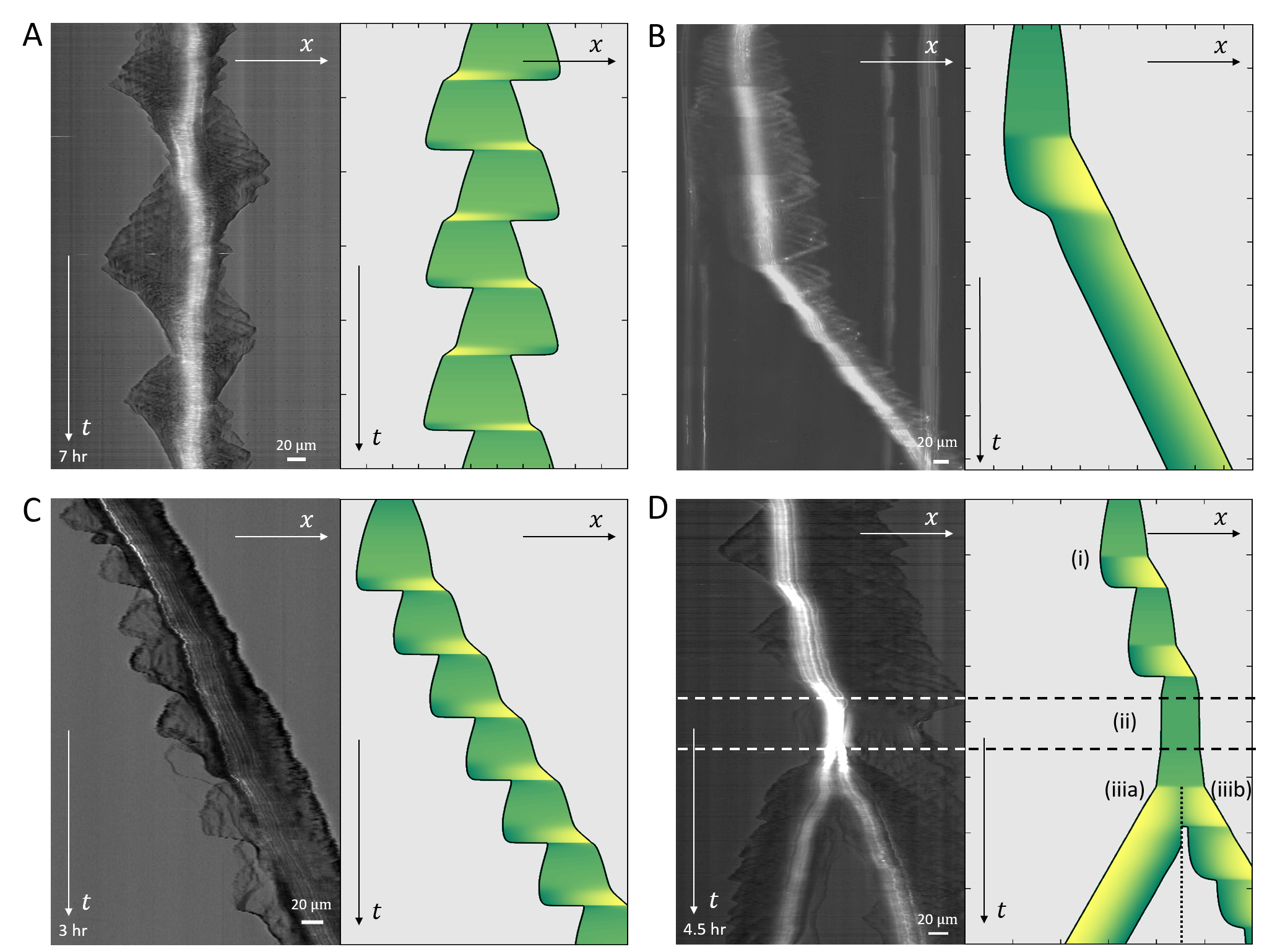}
\caption{\label{fig:expts_02} Comparison of different migration patterns between experiments and model. (A,C,D) Kymographs of C6 glioma cells migrating on laminin-coated lines of 5$\mu$m width (frames are 30 sec apart). A) Stick-slip motion with frequent direction changes. Model parameters: $\beta = 18$, $c=9$, $d=30$, $k=1.05$, $f_s=5$, $r=0.5$, $\kappa=20$, $\delta=220$. B) Kymograph of 3T3 cells seeded on a nanofiber coated with fibronectin (image extracted from \cite{guetta2015protrusive}). The cell initially spreads by developing lamellipodia in both directions, and eventually breaks symmetry and migrates persistently with an approximately constant length. Model parameters: $\beta = 8$, $c = 3.85$, $d = 3.85$, $k=2$, $f_s=5$, $r=1$, $\kappa=20$, $\delta=220$. C) Persistent stick-slip motion, with large lamellipodia extensions at the rear of the cell (image extracted from \cite{monzo2016mechanical}). Model parameters: $\beta=40$, $c=9$, $d=40$, $k=1.1$, $f_s=5$, $r=0.4$ $\kappa = 20$, and $\delta=60$. D) A cell performing stick-slip, goes into mitosis (non-motile), divides and generates two daughter cells migrating in opposite directions. The transitions between the different stages are indicated by the horizontal dashed lines. Model parameters: (i) $\beta=40$, $r=10$ (ii) $\beta=15$, $r=10$ (iiia) $\beta = 19$, $r=12$ (iiib) $\beta = 19$, $r=20$. Rest of the parameters $c=9$, $d=40$, $k=1.1$, $f_s=5$,$\kappa = 20$, $\delta=300$.}
\end{figure}

In Fig.\ref{fig:expts_02} we give a wide range of different cell migrations observed in experiments, demonstrating that the model can produce a similar range of cell migrations. In Fig.\ref{fig:expts_02}A we show a cell that seems to be oscillating between being polarized to the right and to the left, such that it hardly migrates over time. In our model a similar behavior may occur if the cell is in the stick-slip regime, and following each slip event it is shorter than the critical length, allowing for noise to induce direction changes (as explained in Figs.\ref{fig:symV_04},\ref{fig:symV_05}).

In Fig.\ref{fig:expts_02}B a kymograph is shown of a cell that is initially at rest, and elongates symmetrically. At some later time it breaks the right-left symmetry and moves in a persistent manner. This behavior is well described by our model, where the rest-length of the cell is below the polarization length, but the conditions are such that $\beta>\beta_c$ and it elongates symmetrically until it reaches a length that is longer than the critical length $l_c$. At a later time point we decreased the value of the surface adhesion strength $r$, leading to a transition from long-slow to short-fast migration mode,  which seems to fit this cellular behavior (similar to Fig.\ref{fig:constV_05}).

In Fig.\ref{fig:expts_02}C we plot a kymograph of a cell that is persistently moving on a 1D track in the stick-slip regime. During each stick-slip event the progression of the cell's leading edge is slightly modulated, while at the back there are periodic extensions of lamellipodia that undergo large retractions. This behavior corresponds very well to the stick-slip behavior in our model.

Finally, in Fig.\ref{fig:expts_02}D we present a kymograph of a cell that is first observed to be migrating by slow stick-slip events. At some point the cell stops and rounds up, as it enters mitosis, from which two daughter cells emerge, both migrating with similar speeds, but one seems to be moving smoothly while the other exhibits some stick-slip cycles. We demonstrate a similar chain of migration changes using our model, by modulating the parameters that determine the actin treadmilling activity ($\beta$) and adhesion strength ($r$). 

These comparisons serve to show that the model can describe the complex migration patterns of cells. The parameters used in the model may correspond to a unique migration mode, or correspond to regimes where different migration modes coexist. The transitions in the cell behavior from one migration mode to another may therefore correspond to the dynamics of the internal parameters of the cell. In addition, noise in the dynamical variables (such as in the actin treadmilling speed), can drive the enhanced polarization changes observed during stick-slip motion.

\section*{Conclusion}

Cell motility involves a large number of cellular components, connected by a complex network of interactions. In addition, the system is noisy, which makes the analysis of cell migration a daunting task. In the present work we developed a theoretical model aimed at exposing the motility patterns of cells moving along a one-dimensional track. One dimensional motion is both simpler to analyze experimentally, and describe theoretically, as well as highly relevant to cells migrating within tissues during development and cancer progression.

We find that the coupling between the cell length and the slip-bonds of adhesion molecules, through the spring-like elasticity of the cell, provides the basic mechanism of stick-slip during cell migration. Furthermore, we show that when cells migrate and their polarization is maintained by the overall actin treadmilling flow, the cell length couples strongly to the polarization state of the cell: cells below a critical length do not polarize. While persistent smooth and stick-slip migration modes arise naturally in our model, we predict that stick-slip events may allow for noise-induced direction changes, as the cell recoils to less than the critical length at each stick-slip cycle.

The minimal model that we present recovers the rich variety of cellular migration patterns, which we then compare to experiments on migrating cancer cells (glioma). These comparisons validate the model and demonstrate how it can provide the framework for understanding the complex migration patterns exhibited by migrating cells. 

Our model makes detailed predictions regarding the dependence of the cellular migration modes on the coarse-grained parameters of the model, which describe the cell's mechanics ($k$), actin-polymerization activity ($\beta$) and surface adhesion ($r$). We predict a reentrant smooth migration regime as function of the surface adhesion, with stick-slip occurring only at the intermediate regime (Figs.\ref{fig:constV_02},\ref{fig:dynamicV_02},\ref{fig:dynamicV_04}).  

Despite the fact that cell motility is very noisy, our models help expose the underlying \emph{deterministic patterns} of motility \cite{lavi2016deterministic} that drive the cellular motion, which may be partially masked by noise. We show that the noise can also drive dramatic transitions between the different motility patterns, since the cell may reside close to the transition lines, or in a regime of coexistence, between such modes. These results ofer a new explanation for the large "phenotypic heterogeneity" (CCV) that is observed in cell migration experiments, even under well controlled conditions and monoclonal cell population. These results offer a new framework to explain experimental observations of migrating cells, resulting from noisy switching between underlying deterministic migration modes. Small fluctuations in the cellular components that are greatly amplified due to shifting the cells' internal state between coexisting migration modes, and across phase transition lines. The richness in migration modes that our model predicts for cells with identical or slightly different internal states offers a new paradigm to explain "phenotypic heterogeneity" in cell migration.

Our results explain naturally many observations of cells that grow symmetrically, and migrate along 1D tracks. In addition, the model makes predictions that can motivate future experimental exploration. Finally, having a model for single-cell motility can serve as the basis for the description of collective cell migration \cite{jain2019emergence}, which goes beyond treatments of the cells as simplified self-propelled particles.

\begin{acknowledgments}
NSG acknowledges that this work is made possible through the historic generosity of the Perlman family. NSG is the incumbent of the Lee
and William Abramowitz Professorial Chair of Biophysics and this research was supported by the Israel Science Foundation
(Grant No. 1459/17). NCG and PM acknowledge support by IFOM starting package, and the Italian Association for Cancer Research (AIRC), Investigator Grant (IG) 20716 to NCG.
\end{acknowledgments}


\appendix

\section{Analysis of the stick-slip phase diagram as function of the adhesion saturation parameter $r_0$.}
\begin{figure}[h]
\includegraphics[width=\linewidth]{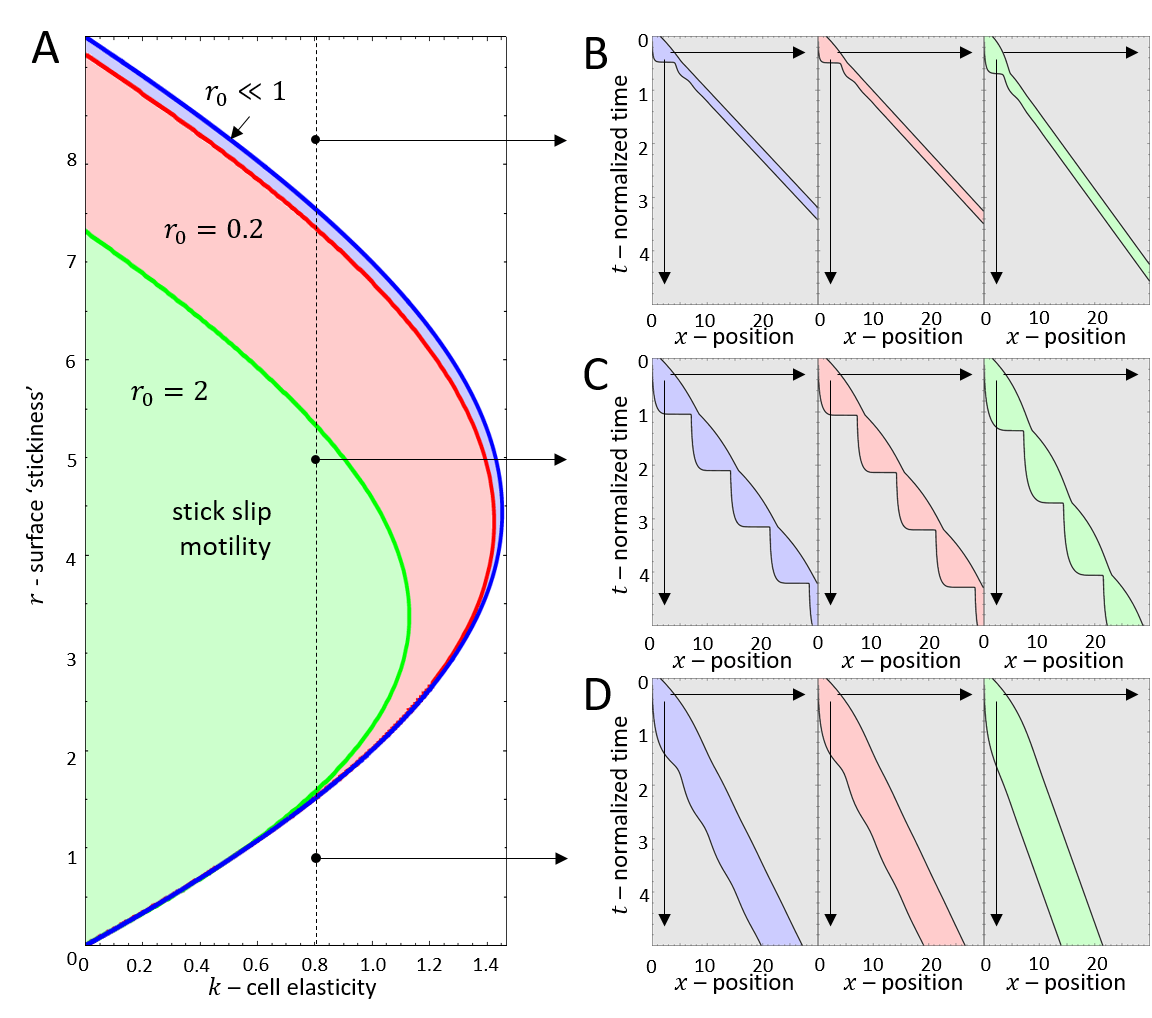} 
\caption{\label{fig:SI_02} The effects of the adhesion saturation parameter $r_0$ on the cellular dynamics. A) $k-r$ phase diagram for different values of $r_0$. Blue/Red/Blue curves represent the transition between migration in constant length and stick slip for. A-D) kymographs of the model for $k=0.8$ and $r=2,5,8.25$. Blue/Red/Green colors represent $r_0\rightarrow 0$/$r_0=0.2$/$r_0=2$.} 
\end{figure}

In Eqs.\ref{Fp1},\ref{Fp2} the parameters $\alpha$ and $\gamma$ should be multiplied by a factor of $r/(r+r_0)$, but throughout the main text (Fig.\ref{fig:constV_02}) we treated the limit of $r_0\rightarrow 0$. Here we show the effects of a finite value of $r_0$. Fig. \ref{fig:SI_02}  demonstrates that as the value of $r_0$ increases the stick-slip region along the $k-r$ phase diagram decreases, and the cells migrate with a slower speed.


\section{Concentration profile of the polymerization inhibitor $c(x)$}
Consider a generic polarity cue which follows an advection diffusion transport along the length of the cell $l$. The polarity cue acts on both ends of the cell, and inhibits actin polymerization.

The derivation starts by dividing the cue into two populations: 1) bound cue proteins $c_b$ which are advected by the retrograde flow $v$,  and 2) unbound cue proteins $c_f$ which diffuse with a diffusion coefficient $D$.

The bound/unbound proteins can attach/detach to/from the advected actin, with rates $\bar{k}_{on}/\bar{k}_{off}$ respectively. The total concentration of the polarity cue in the cell is considered as conserved, i.e, $c_{tot}=c_b(x,t)+c_f(x,t)$ $\forall$ $\lbrace t,x\rbrace$.

The advection-diffusion transport equations of $c_b$ and $c_f$ are given by
\begin{eqnarray} \label{cb}
\frac{\partial c_b(x,t)}{\partial t} &=& v\frac{\partial c_b(x,t)}{\partial x}+\bar{k}_{on} c_f(x,t)-\bar{k}_{off}c_b(x,t)\\ \label{cf}
\frac{\partial c_f(x,t)}{\partial t} &=& D\frac{\partial^2 c_f(x,t)}{\partial x^2}-\bar{k}_{on} c_f(x,t)+\bar{k}_{off}c_b(x,t) \nonumber \\
\end{eqnarray}

At the limit of fast exchange, in which the kinetic rates $\bar{k}_{on},\bar{k}_{off}\gg \frac{D}{\delta x^2},\frac{v}{\delta x}$, along any segment $\delta x\in[0,L]$ , (Eqs.\ref{cb},\ref{cf})  reduce to
\begin{equation} \label{ss}
\frac{\partial c(x,t)}{\partial t}\left(1+\frac{\bar{k}_{on}}{\bar{k}_{off}}\right)=\frac{\partial}{\partial x}\left(v c(x,t)+\frac{\bar{k}_{on}}{\bar{k}_{off}}D\frac{\partial c(x,t)}{\partial x}\right)
\end{equation}
where $c(x,t)\equiv c_b(x,t)$.

At steady state, (Eq.\ref{ss}) takes the form of 
\begin{equation} \label{diff}
0=\frac{\partial}{\partial x}\left(v\space c(x)+D\frac{\partial c(x,t)}{\partial x}\right)
\end{equation} 
where $\frac{\bar{k}_{on}}{\bar{k}_{off}}D \rightarrow D$.

The solution of (Eq.\ref{diff}) reads
\begin{equation}
c(x)=c_0e^{-\frac{v x}{D}}+c_1
\end{equation}

and the coefficients are obtained by applying a no-flux boundary condition at $x_f$ and $x_b$ 
\begin{equation}
0=v c(x) +D\left.\frac{\partial c(x)}{\partial x}\right|_{x=x_f,x_b}\rightarrow c_1=0
\end{equation}
By considering mass conservation we obtain
\begin{equation}
c_{tot}=c_0\int_{x_b}^{x_f} e^{-\frac{vx}{D}}dx\rightarrow c_0=\frac{c_{tot}v}{D}\left(\frac{1}{e^{{-\frac{v x_b}{D}}}-e^{-\frac{v x_f}{D}}}\right)
\end{equation}
and therefore the concentration profile is given by Eq.\ref{cx}.

\section{Rescaling the velocity equation in the asymmetric self-polarized model}
\begin{eqnarray} \label{vss0}
&&v_{ss}\left(x_f,x_b\right)=\beta\left(\tilde{c}(x_f)-\tilde{c}(x_b)\right)
=\beta\left(\frac{c_s}{c_s+c(x_f)}-\frac{c_s}{c_s+c(x_b)}\right)\nonumber\\
=&&\beta\left(\frac{1}{1+\frac{c_{tot}}{c_s}\frac{v}{D}\left(\frac{e^{-\frac{v x_f}{D}}}{e^{{-\frac{v x_b}{D}}}-e^{-\frac{v x_f}{D}}}\right)}-\frac{1}{1+\frac{c_{tot}}{c_s}\frac{v}{D}\left(\frac{e^{-\frac{v x_b}{D}}}{e^{{-\frac{v x_b}{D}}}-e^{-\frac{v x_f}{D}}}\right)}\right) \nonumber \\
\end{eqnarray}

Next, we rescale Eq.(\ref{vss}) by the time and length scale $k_{off}^{-1}$ and $x_0$, such that $\frac{x_0^{2}}{k_{off}^{-1}}D\rightarrow D, \frac{k_{off}^{-1}v}{x_0}\rightarrow v$, and $\frac{k_{off}^{-1}\beta}{x_0}\rightarrow \beta$ $\frac{c_{tot}x_0}{c_{s}}\rightarrow c$, 
and then rescale $x_0x_f\rightarrow x_f$, $x_0x_b\rightarrow x_b$ and $\frac{c_{tot}}{c_{s}x_0}\rightarrow c$ which gives
\begin{eqnarray}
&&v_{ss}(x_f,x_b)=\nonumber\\
=&&\beta\left(\frac{1}{1+c\frac{v}{D}\left(\frac{e^{-\frac{v x_f}{D}}}{e^{{-\frac{v x_b}{D}}}-e^{-\frac{v x_f}{D}}}\right)}-\frac{1}{1+c\frac{v}{D}\left(\frac{e^{-\frac{v x_b}{D}}}{e^{-\frac{v x_b}{D}}-e^{-\frac{v x_f}{D}}}\right)}\right)\nonumber\\
&&=\beta\left(\frac{1}{1+c\frac{v}{D}\left(\frac{1}{e^{{\frac{v (x_f-x_b)}{D}}}-1}\right)}-\frac{1}{1+c\frac{v}{D}\left(\frac{1}{1-e^{-\frac{v (x_f-x_b)}{D}}}\right)}\right) \nonumber\\
\end{eqnarray}
and finally we change coordinates to $l=x_{f}-x_{b}$, and obtain Eq.\ref{vss2}.

\section{Experimental methods}
\subsection{Cell culture} C6 rat glioma cells were obtained from ATCC and grown in High Glucose DMEM supplemented with 10 \% heat inactivated FBS (HI-FBS) and glutamine (Invitrogen). Graded brain tumor specimens were obtained with informed consent, as part of a study protocol approved by the SingHealth Centralised Institutional Review Board A (Singapore). From one of those specimen, derived human glioma propagating cells NNI21 (generous gift from Carol Tang lab, National Neuroscience Institutte Singapore were isolated and cultured as described previously \cite{chong2009cryopreservation} as tumor spheres in high glucose DMEM/F12 (1:1) supplemented with sodium pyruvate, non-essential amino acid, penicillin/streptomycin, glutamine, B27 supplement (Invitrogen), bFGF (20ng/ml), EGF (20ng/ml) (PeproTech), and heparin (5µg/ml) (Sigma). For transfection and migration assays, NNI21 were cultured as monolayers on laminin (10 µg/ml) coated petri dishes for 3-5 days before transfection. HGPCs transfections were performed with a Neon electroporator (Invitrogen) as per manufacturer’s recommendations using vinculin-mCherry (gift from P. Kanchanawong, Mechanobiology Institute, National University of Singapore, Singapore) as an adhesion rapporteur and GFP-Plasma Membrane (GFP-PM, Clonetech) for plasma membrane staining.

\subsection{Micropatterning} 5$\mu$m lines were printed on glass coverslips using deep UV photopatterning technique as directed in \cite{azioune2010protein}. Briefly, 25-mm glass coverslips were plasma-treated for 5 min and incubated for 1h at room temperature (RT) with poly-l-lysine–grafted–polyethylene glycol (0.1 mg/ml, pLL-PEG, SuSoS) diluted in Hepes [10 mM (pH 7.4)]. After washing in water, the pLL-PEG–covered coverslip was placed with the polymer brush facing downward onto the chrome side of a quartz photomask for photolithography treatment (7-min ultraviolet-light exposure). Subsequently, the coverslip was removed from the mask and coated with laminin (10 $\mu$g/ml) (Invitrogen) diluted in dPBS for 1h at 37C. Cells were seeded on the patterns and incubated at 37 C. Cell imaging typically started within the following hour.
\subsection{Microscopy} Phase contrast of live specimens were performed on a Leica AM TIRF MC system equipped with temperature, humidity, and CO2 control. Long term imaging was done using a 10X objective (Leica HCX PL FLUOTAR 10x/0.30NA PH1 Objective). Acquisitions were typically obtained over a period varying from 2 to 12 h (1 image/30 s). 
\section{Bifurcation analysis}
\begin{figure*}[htbp!]
\includegraphics[width=\linewidth]{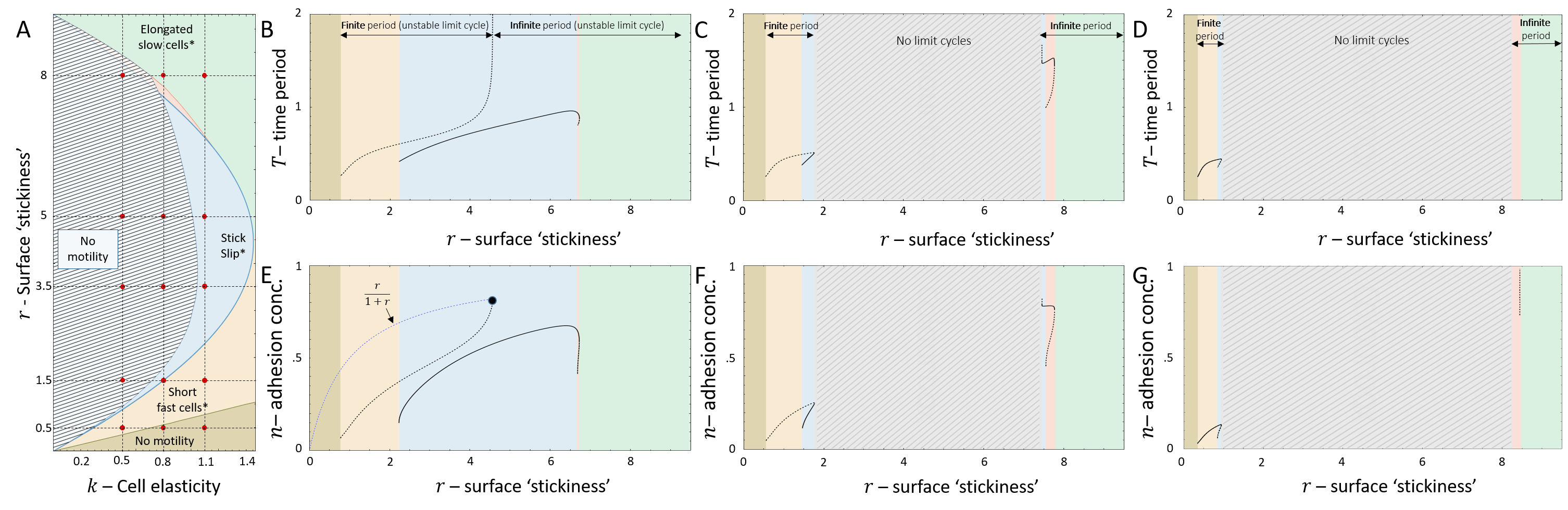} \caption{\label{fig:SI_09} A) $k-r$ phase diagram as presented in Fig. \ref{fig:dynamicV_03}. Black dashed lines correspond to the $r$ cross section for $k=0.5,0.8,1.1$ and $k$ cross sections of $r={0.5,1.5,3.5,5,8}$. Red dots correspond to intersections (dynamics at the intersections are shown in Figs. \ref{fig:SI_10},\ref{fig:SI_11},\ref{fig:SI_12}). B-C) The time period of the limit cycles in the model along the cross sections of $k=1.1$/$0.8$/$0.5$ respectively. Black dashed line indicate the unstable limit cycles. Black solid lines indicate the stable limit cycles. E-G) Maximal amplitude of the $n$ state variable as a function of $r$ for the limit cycles in the vector field along the cross section of$k=1.1$/$0.8$/$0.5$ respectively. Black dashed curves correspond to the unstable limit cycles. Solid Black curves correspond to the stable limit cycles. Dashed blue curve corresponds to the $n$ value of the fixed point in the region where $l<x_0$ ($l<1$). Black dot corresponds to the transition between a finite and infinite period of the discontinuous unstable limit cycle (the $n$ cross section in which the unstable limit cycle crosses the section where $n=\frac{r}{1+r}$.
 Parameters: $f_s=5$,$\kappa=20$,$\beta=11$, $c=3.85$, $d=3.85$.} 
\end{figure*}

\begin{figure*}[h]
\includegraphics[width=\linewidth]{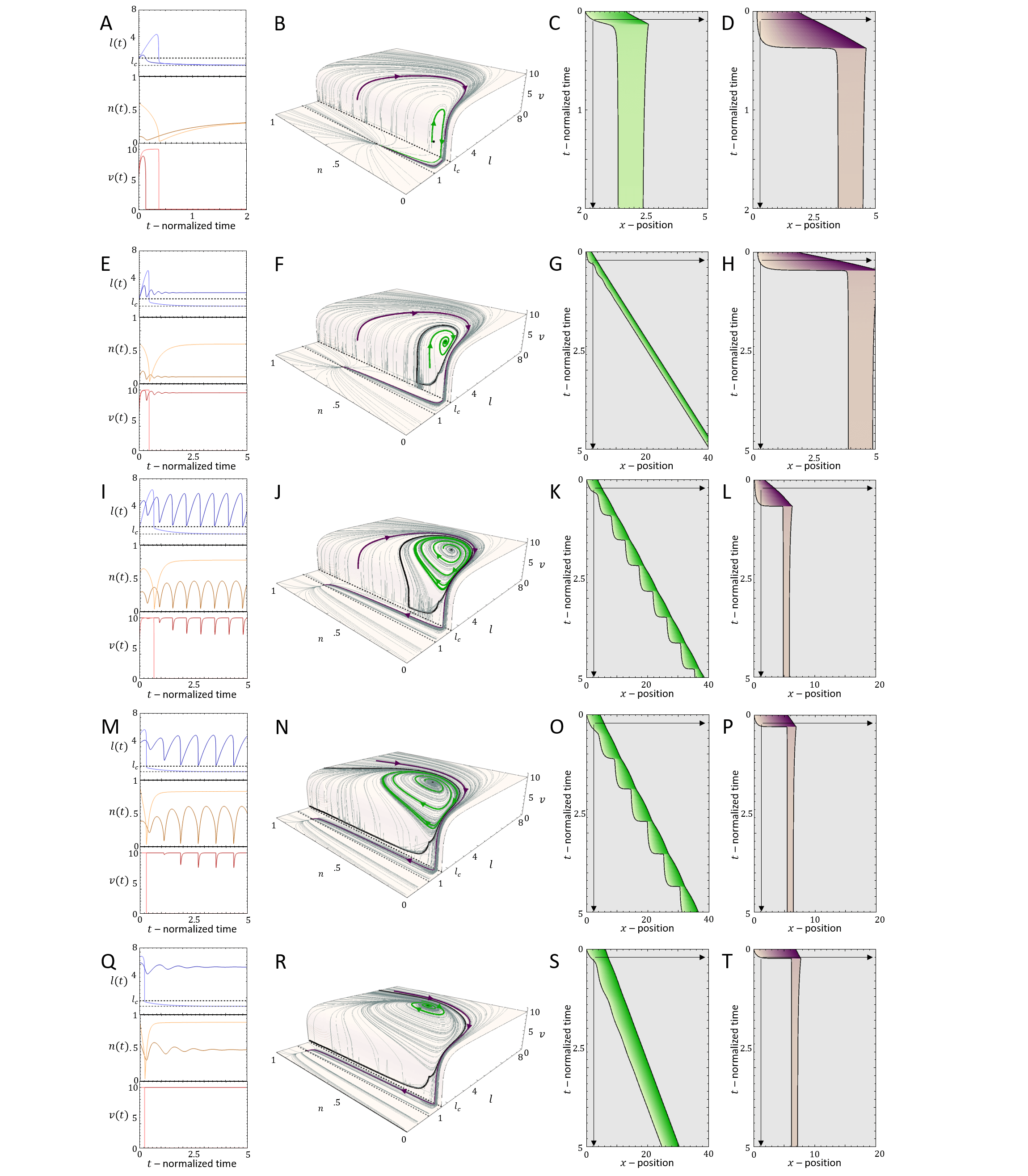} 
\caption{\label{fig:SI_10} The dynamics along the line of constant $k=1.1$ (vertical dashed line in Fig.\ref{fig:SI_09}A). A-D) $r=0.5$. E-H) $r=1.5$. I-L) $r=3.5$. M-P) $r=5$. Q-T) $r=8$. Blue/Orange/Red curves at panels A,E,I,M,Q correspond to the time series of the cell length, adhesion concentration and actin retrograde flow respectively (bold/thin lines corresponds to green/purple trajectory). Green and purple curves on panels B,F,J,N,R demonstrate the trajectories in the $l-n-v$ phase space. Black solid curves are the separatrices. Panels D,G,I display the corresponding kymographs. Parameters: $f_s=5,\kappa=20,\beta=11$,$c=3.85$,$d=3.85$.} 
\end{figure*}

\begin{figure*}[h]
\includegraphics[width=\linewidth]{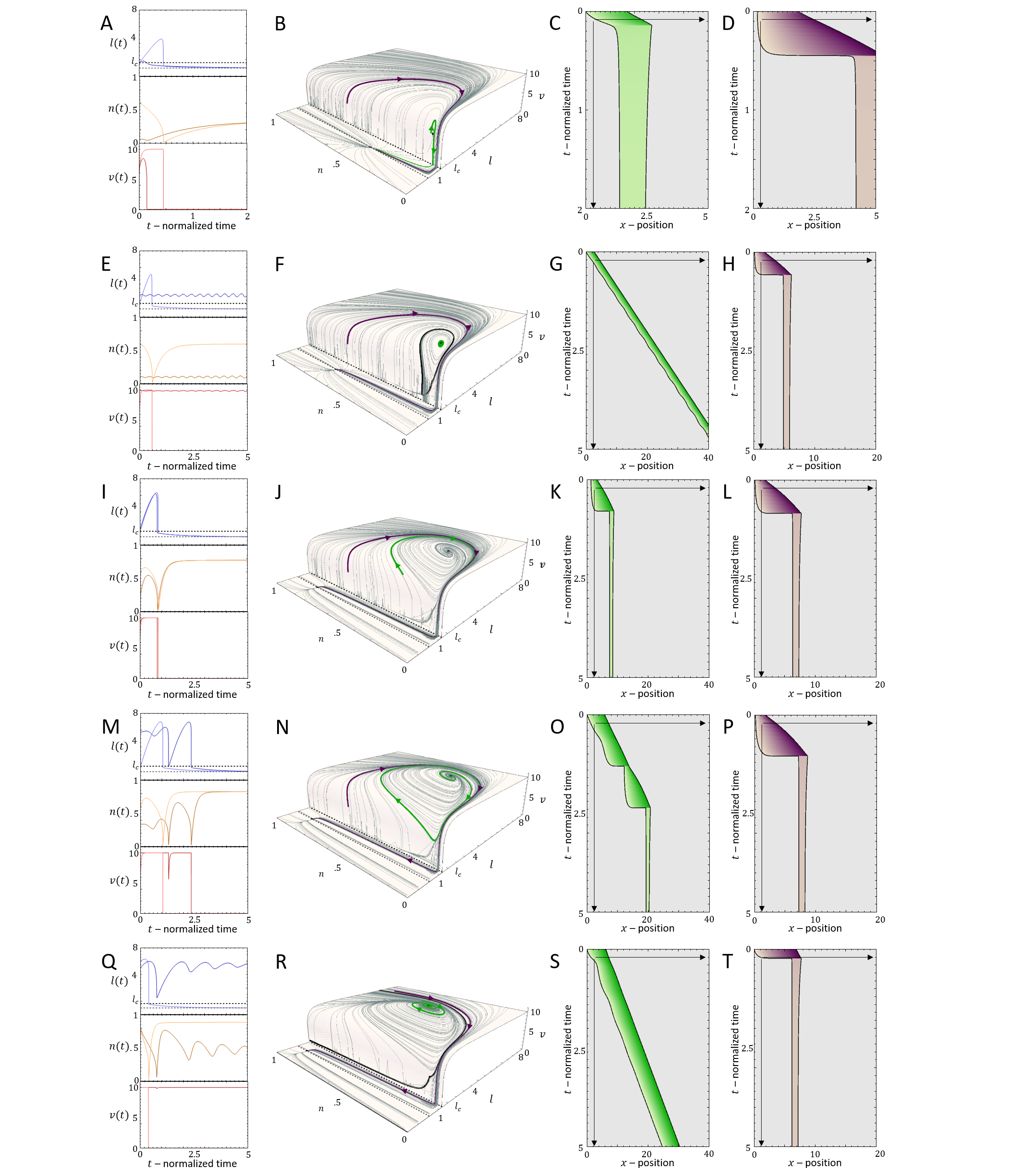} 
\caption{\label{fig:SI_11} The dynamics along the line of constant $k=0.8$ (vertical dashed line in Fig.\ref{fig:SI_09}A). A-D) $r=0.5$. E-H) $r=1.5$. I-L) $r=3.5$. M-P) $r=5$. Q-T) $r=8$. Blue/Orange/Red curves at panels A,E,I,M,Q correspond to the time series of the cell length, adhesion concentration and actin retrograde flow respectively (bold/thin lines corresponds to green/purple trajectory). Green and purple curves on panels B,F,J,N,R demonstrate the trajectories in the $l-n-v$ phase space. Black solid curves are the separatrices. Panels D,G,I display the corresponding kymographs. Parameters: $f_s=5,\kappa=20,\beta=11$,$c=3.85$,$d=3.85$.} 
\end{figure*}

\begin{figure*}[h]
\includegraphics[width=\linewidth]{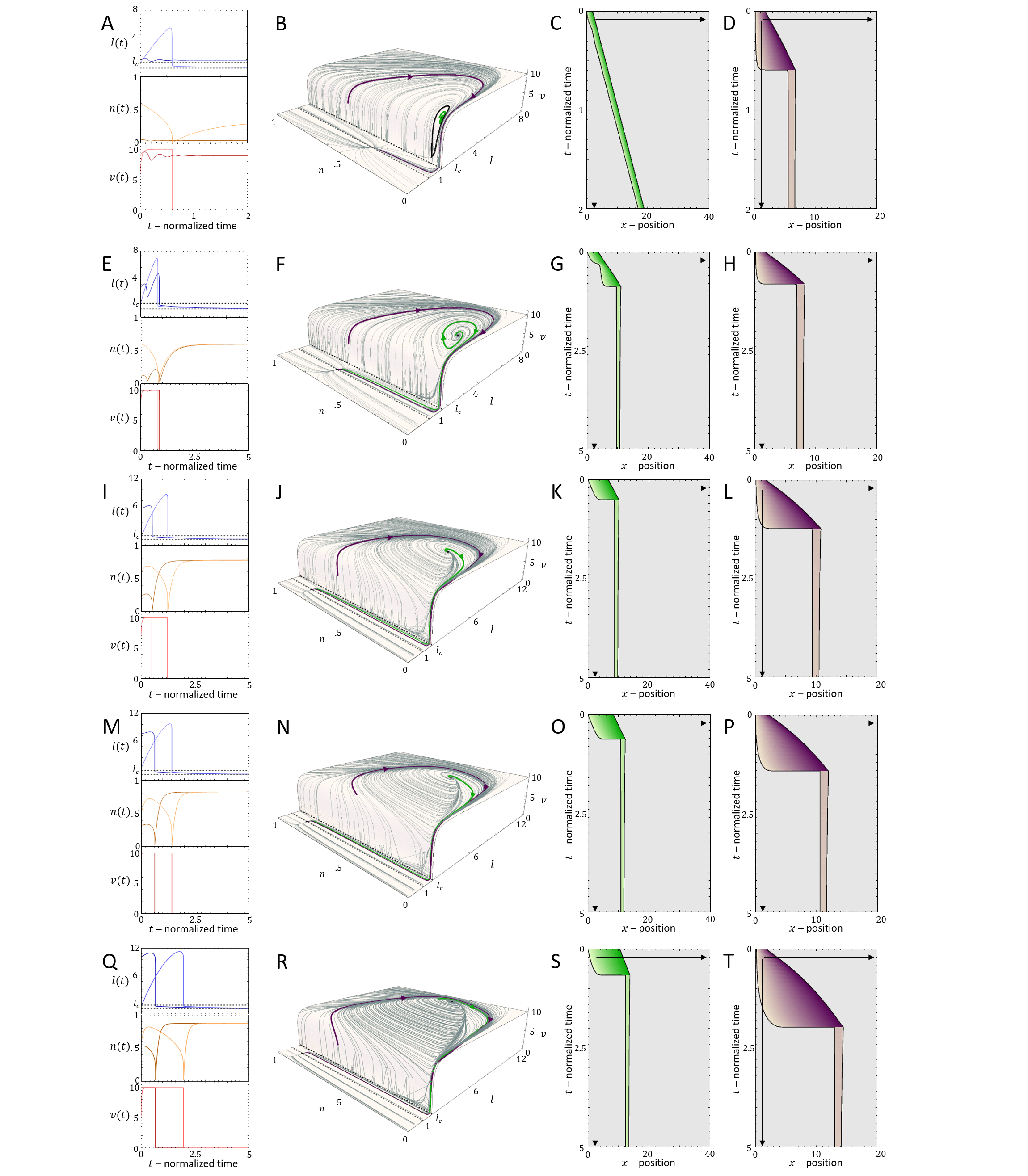} 
\caption{\label{fig:SI_12} The dynamics along the line of constant $k=0.5$ (vertical dashed line in Fig.\ref{fig:SI_09}A). A-D) $r=0.5$. E-H) $r=1.5$. I-L) $r=3.5$. M-P) $r=5$. Q-T) $r=8$. Blue/Orange/Red curves at panels A,E,I,M,Q correspond to the time series of the cell length, adhesion concentration and actin retrograde flow respectively (bold/thin lines corresponds to green/purple trajectory). Green and purple curves on panels B,F,J,N,R demonstrate the trajectories in the $l-n-v$ phase space. Black solid curves are the separatrices. Panels D,G,I display the corresponding kymographs. Parameters: $f_s=5,\kappa=20,\beta=11$,$c=3.85$,$d=3.85$.} 
\end{figure*}

The introduction of the actin dynamics in Part 2, introduces to the $n-l$ vector field a discontinuity at $l=l_c$. The vector discontinuity gives rise  to an additional unstable limit cycle due to a trajectory that collides with the line of $l=l_c$  (Figs.\ref{fig:dynamicV_02},\ref{fig:dynamicV_03}). The period of the limit cycle is finite as long as the amplitude of the $n$ coordinate is limited to values: $n<\frac{r}{1+r}$. When the $n$ coordinate amplitude grows beyond this limit, $n>\frac{r}{1+r}$, the period of the discontinuous unstable limit cycle is infinite (Figs. \ref{fig:SI_09},\ref{fig:SI_10},\ref{fig:SI_11},\ref{fig:SI_12}).



\bibliography{ref}

\end{document}